\documentclass[12pt, oneside]{article}   	
\usepackage{geometry}                		
\usepackage{graphicx}				
\usepackage{amssymb}
\usepackage{subfig}
\usepackage{amsmath}
\usepackage{color}
\usepackage{graphicx}
\usepackage{booktabs,caption}
\usepackage{setspace}
\usepackage{natbib}
\setcitestyle{authoryear,open={(},close={)}}
\usepackage[utf8]{inputenc}
\usepackage[english]{babel}
\usepackage{enumitem}

\usepackage{tikz}
\usetikzlibrary{arrows,shapes.arrows,shapes.geometric,backgrounds,decorations.pathmorphing,positioning,fit,automata}

\tikzset{
	>=stealth',
	true/.style={
		rectangle,
		draw=black, very thick,
		text width=6.5em,
		minimum height=2em,
		text centered,
		fill=gray, opacity = 0.5},
	punkt/.style={
		rectangle,
		rounded corners,
		draw=black, very thick,
		text width=6.5em,
		minimum height=2em,
		text centered},
	est/.style={
		circle,
		draw=black, very thick,
		text centered},
	shade/.style={
		circle,
		draw=black, very thick, fill=gray!50,
		text centered},
	weight/.style={
		circle,
		draw=black, very thick,
		text width=6.5em,
		minimum height=2em,
		text centered},
	pil/.style={
		->,
		thick,
		shorten <=2pt,
		shorten >=2pt,},
	double/.style={
		<->,
		thick,
		shorten <=2pt,
		shorten >=2pt,},
	dash/.style={
		dashed,
		thick,
		shorten <=2pt,
		shorten >=2pt,},
	dashdouble/.style={
		<->,
		dashed,
		thick,
		shorten <=2pt,
		shorten >=2pt}}

\newtheorem{theorem}{Theorem}
\newtheorem{lemma}[theorem]{Lemma}

\date{}							
\title{Robust inference on population indirect causal effects: the generalized front-door criterion}
\begin{document}

\maketitle
\noindent Isabel R. Fulcher \\
\noindent \textit{Department of Biostatistics, Harvard T.H. Chan School of Public Health, Boston, USA} \\

\noindent Ilya Shpitser\\
\noindent \textit{Department of Computer Science, Johns Hopkins University, Baltimore, USA} \\

\noindent Stella Marealle \\
\noindent \textit{D-tree International, Zanzibar, Tanzania} \\

\noindent Eric J. Tchetgen Tchetgen \\
\noindent \textit{Wharton Statistics Department, University of Pennsylvania, Philadelphia, USA} \\
	
		\onehalfspacing
\begin{abstract}
	Standard methods for inference about direct and indirect effects require stringent no unmeasured confounding assumptions which often fail to hold in practice, particularly in observational studies. The goal of this paper is to introduce a new form of indirect effect, the population intervention indirect effect (PIIE), that can be nonparametrically identified in the presence of an unmeasured common cause of exposure and outcome. This new type of indirect effect captures the extent to which the effect of exposure is mediated by an intermediate variable under an intervention that holds the component of exposure directly influencing the outcome at its observed value. The PIIE is in fact the indirect component of the population intervention effect, introduced by Hubbard and Van der Laan (2008). Interestingly, our identification criterion generalizes Judea Pearl's front-door criterion as it does not require no direct effect of exposure not mediated by the intermediate variable. For inference, we develop both parametric and semiparametric methods, including a novel doubly robust semiparametric locally efficient estimator, that perform very well in simulation studies. Finally, the proposed methods are used to measure the effectiveness of monetary saving recommendations among women enrolled in a maternal health program in Tanzania.
\end{abstract}


\section{Introduction} 

The population average causal effect is by in large the most common form of total effect evaluated in observational data due to the natural connection to scientific queries arising from randomized studies. However, alternate forms of total effect may be of greater interest in observational studies with harmful exposure such that one may not want to conceive of a hypothetical intervention that forces a person to be exposed. \cite{hubbard2008population} define the population intervention effect (PIE) of an exposure as the contrast relating the mean of an outcome in the population to that in the same observed population had no one been exposed. Interestingly, the PIE is closely related to the effect of treatment on the treated (ETT) and attributable fraction (AF), which have also been toted as causal quantities to assess public health impact of a harmful exposure \citep{geneletti2007defining,hahn1998role, sjolander2010doubly, greenland1993maximum}. The ETT compares the outcome among those exposed to the potential outcome had they not been exposed -- for binary treatment, the PIE is equal to the ETT scaled by prevalence of treated persons. The AF is the proportion of potential outcome events that would be eliminated from the observed population had contrary to fact no one been exposed -- for binary outcome, the PIE is equal to the AF scaled by prevalence of outcome. As such, the PIE is a scale-dependent version of these quantities and may be of greater interest when evaluating the potential impact of programs that eliminate a harmful exposure from a population.

Recent causal mediation methods have been developed to decompose such total causal effects into direct and indirect pathways through a mediating variable \citep{pearl2001direct,vansteelandt2012natural, sjolander7mediation}.  Although the natural (pure) direct and indirect effects of the average causal effect (ACE) are the most common form of mediated causal effects, researchers have argued that the direct and indirect components of the ETT and AF are equally of scientific interest and may in fact require weaker conditions for identification \citep{vansteelandt2012natural}. Namely, identification of natural direct and indirect effects requires the stringent assumption that there is no unmeasured confounding of the exposure-outcome, exposure-mediator, and mediator-outcome associations and no exposure induced confounding of the mediator-outcome association, even by measured factors \citep{pearl2001direct, avin2005identifiability}. \cite{vansteelandt2012natural} propose a particular form of direct and indirect effects of the ETT which they show remain identified in the presence of exposure-mediator unmeasured confounding. This is an important result for settings where a randomized experiment is impractical or unethical such that observational data must be used and unmeasured confounding effects of the exposure cannot be ruled out. Unfortunately, \cite{vansteelandt2012natural} are unable to identify the indirect effect whenever the exposure-outcome association is confounded. In this paper, we propose an alternative form of indirect effect and describe sufficient conditions for nonparametric identification in the presence of unmeasured confounding of the exposure-outcome association, therefore complementing the results of \cite{vansteelandt2012natural}.


Specifically, we propose a decomposition of the PIE into the population intervention direct effect (PIDE) and population intervention indirect effect (PIIE). The PIIE is interpreted as the contrast between the observed outcome mean for the population and the population outcome mean had contrary to fact the mediator taken the value it would have in the absence of exposure. Thus, the PIIE is relative to the current distribution of an exposure and does not require conceiving of an intervention that would force an individual to take a harmful level of exposure in the case of binary exposure \citep{hubbard2008population}. Our approach leads to an alternative effect decomposition of the ETT and AF of \cite{vansteelandt2012natural} and \cite{sjolander7mediation} (up to a scaling factor). Notably, we establish that the PIIE can be identified even when there is an unmeasured common cause of exposure and outcome variables, provided it is not also a cause of the mediator. This estimand may be of interest in a variety of settings where unmeasured confounding of the exposure-outcome relation cannot be ruled out with certainty. For example, in recommender systems, the assignment mechanism for the mediator (e.g. recommendation) is typically known or under control of the researcher, such that unmeasured confounding of the exposure-mediator and mediator-outcome relations are not of concern. The application considered in this paper investigates the indirect effect of a woman's pregnancy risk on monetary savings for delivery mediated by the amount she is recommended to save by a community health worker. Note that the PIDE does not share this identification result and is identified under the same conditions as the natural direct effect. 

Beyond its inherent scientific interest as quantifying the mediated component of the PIE, the PIIE may also be viewed as an approach to partially identify a total effect of an exposure on an outcome subject to unmeasured confounding in settings where one might be primarily interested in such a total effect. Interestingly, the identifying formula we obtain for the PIIE matches Judea Pearl's celebrated front-door formula, a well-known result for identification of the total effect in the presence of unmeasured confounding given that (1) a mediating variable(s) intercepts all directed paths from exposure to outcome so that the indirect effect equals the total effect and (2) there is no unmeasured confounding of the mediator-outcome or exposure-mediator associations \citep{pearl2009causality}. In the setting where an investigator believes they have captured one or more mediating variables that satisfy the front-door criterion, they can use our proposed methodology to estimate either the PIE or the average causal effect. Notably, identification of indirect effects with Pearl's front-door criterion requires a key assumption of no direct effect of the exposure on the outcome not through the mediator in view. In contrast, our generalized front-door criterion allows for presence of such direct effects. Thus, even if an investigator cannot satisfy  criteria (1), they may still be able to capture the un-confounded component of the PIE through one or more mediating variables. Compared to other methods that relax the assumption of no unmeasured confounding to identify causal effects, our approach applies more generally as it does not require a valid instrumental variable, measuring one or more negative control variables, or parametric assumptions for identification \citep{angrist1996identification,imbens2008regression,campbell2015experimental,lipsitch2010negative,miao2017invited,vanderweele2011bias}. We emphasize that while the front-door criterion has long been established, the proposed generalized front-door criterion is entirely new to the literature. In addition to new identification results, we also develop both parametric and semiparametric theory for inference about the PIIE. To the best of our knowledge, the proposed methodology also delivers the first doubly robust estimator of Pearl's front-door formula in the literature.

The rest of the paper is organized as follows, in section 2, we discuss nonparametric identification of the PIIE and PIDE. In section 3, we derive both parametric and semiparametric estimators, including a doubly robust semiparametric locally efficient estimator for the PIIE and PIDE. In section 4, the performance of these estimators is evaluated in a range of settings in extensive simulation studies. In section 5, the proposed methods are used to measure the effectiveness of monetary savings recommendations for delivery among pregnant women enrolled in a maternal health program in Zanzibar, Tanzania.

\section{Nonparametric Identification} 

In the following, let $Z(a)$ denote the counterfactual mediator variable had the exposure taken value $a$ and $Y(a) = Y(a, Z(a))$ denote the counterfactual outcome had exposure possibly contrary to the fact taken value $a$. We will also consider the counterfactual outcome $Y(A,Z(a^*)) = Y(Z(a^*))$ had exposure taken its natural level and the mediator variable taken the value it would have under $a^*$. Note that when $a^* = 0$, $Y(Z(0))$ is the counterfactual outcome had exposure taken its natural level and the mediator variable taken the value it would have under no exposure. Additionally, let $C$ be a set of observed pre-exposure covariates known to confound $A$-$Z$, $A$-$Y$ and $Z$-$Y$ associations. Throughout $Z$ can be vector valued.

We first consider the standard decomposition of the average causal effect (ACE). For exposure levels $a$ and $a^*$, 
\begin{align*}
ACE(a,a^*) & = E[Y(a,Z(a)) - Y(a^*,Z(a^*))] \\
& =  \underbrace{E[Y(a,Z(a)) - Y(a, Z(a^*))]}_\text{Natural Indirect Effect} + \underbrace{E[Y(a, Z(a^*)) - Y(a^*,Z(a^*))]}_\text{Natural Direct Effect}
\end{align*}
The natural indirect effect is the difference between the potential outcome under exposure value $a$ and the potential outcome had exposure taken value $a$ but the mediator variable had taken the value it would have under $a^*$;
$$NIE(a,a^*) = E[Y(a, Z(a)) - Y(a,Z(a^*))]$$ 
The natural direct effect is therefore given by $ACE(a,a^*) - NIE(a,a^*)$.
The NIE and NDE are well-known to be identified under the following conditions \citep{pearl2012causal,imai2010general}:

\begin{align*} 
\textrm{M1.} \ & \textrm{Consistency assumptions: } \textrm{(1) If $A=a$, then $Z(a) =Z$ w.p.1}, \\
& \hspace{4.8cm} \textrm{(2) If $A=a$, then $Y(a) =Y$ w.p.1}, \\
& \hspace{4.8cm} \textrm{(3) If $A=a$ and $Z=z$, then $Y(a,z) =Y$ w.p.1} \\
\textrm{M2.}  \ & Z(a^*) \perp A \mid C=c \ \ \ \forall \ a^*, c \\ 
\textrm{M3.} \ & Y(a,z) \perp  Z(a^*) \mid A=a,C=c \ \ \ \forall \ z,a,a^*,c \\
\textrm{M4.} \ & Y(a,z) \perp  A \mid C=c \ \ \ \forall \ z,a, c
\end{align*}
M1 states the observed outcome is equal to the counterfactual outcome corresponding to the observed treatment. The remaining assumptions essentially state that there is no unmeasured confounding of the exposure and the mediator variable (M2), the mediator variable and the outcome (M3), and the exposure and the outcome (M4). In addition, M3 rules out exposure-induced mediator-outcome confounding. These assumptions could equivalently be formulated under a Nonparametric Structural Equation Model with Independent Errors (NPSEM-IE) interpretation of the diagram in Figure 1a \citep{pearl2009causality}. In addition, define the following positivity assumptions, 
\begin{align*}
\textrm{P1.} & \textrm{ There exists } m_1 >0 \textrm{ such that } f(Z | A, C)  > m_1 \textrm{ almost surely} \\
\textrm{P2.} & \textrm{ There exists } m_2 >0 \textrm{ such that } f(A | C)  > m_2 \textrm{ almost surely}
\end{align*} 

\noindent where $f(Z | A, C)$ and $f(A | C)$ are the probability density functions for $Z|A,C$ and $A|C$, respectively. Under M1-4 and the positivity conditions P1-2, 
\begin{align}
E[Y(a,Z(a^*))] & = \sum_{c,z} E(Y \mid A=a, Z=z, C=c) Pr(Z = z \mid A=a^*, C=c) Pr(C=c) \label{mediation2}
\end{align}
The NIE and NDE fail to be nonparametrically identified if any of assumptions M1-4 fail to hold without an additional assumption \citep{imai2010identification,shpitser2013counterfactual}. 

We will now formally define the decomposition of the population intervention effect under exposure value $a^*$, 
\begin{align}
PIE(a^*) & = E[Y(A,Z(A)) - Y(a^*)]\notag  \\
& = \underbrace{ E[Y(A,Z(A)) - Y(A,Z(a^*)] }_\text{Population Intervention Indirect Effect} + \underbrace{ E[Y(A,Z(a^*)) - Y(a^*,Z(a^*)] }_\text{Population Intervention Direct Effect}  \notag  
\end{align}
The population intervention indirect effect (PIIE) is a novel measure of indirect effect corresponding to the effect of an intervention which changes the mediator from its natural value (i.e. its observed value) to the value it would have had under exposure value $a^*$, 
\begin{align}
PIIE(a^*) = E[Y(A,Z(A)) - Y(A,Z(a^*)] \label{piie}
\end{align}
The PIIE is indeed an indirect effect as it would only be non-null if changing the exposure from its natural value to $a^*$ results in a change in the value of the mediator which in turn results in a change in the value of the outcome. That is, the PIIE captures an effect propagated along the $A \to Z \to Y$ pathway only, and would be null if $A$ has no effect on $Z$ or $Z$ has no effect on $Y$ for all persons in the population. Compared to the NIE, the PIIE only requires intervention on the exposure level of the mediator in the second term and does not require intervention on the exposure level for the potential outcomes for $Y$. Similarly, the Population Intervention Direct Effect (PIDE) is a novel measure of direct effect corresponding to the effect of an intervention which changes the exposure from its natural level to the value under intervention $a^*$, while keeping the mediator variable at the value it would have under intervention $a^*$. This is indeed a direct effect as it would only be non-null if changing the exposure from its natural value to $a^*$, while preventing the mediator variable to change, results in a change in the value of the outcome. That is, the PIDE captures an effect along the $A \to Y$ pathway only. 

The first term of the PIIE, $E(Y)$, is nonparametrically identified; however, the second term requires identification conditions. Identification conditions for the PIIE are less stringent than the NIE as seen by comparing Figure 1a and 1c under a NPSEM-IE interpretation of the diagrams \citep{pearl2009causality}. In fact, the following result states that assumption M4 is no longer needed. 
\begin{lemma}Under assumptions M1-3 and positivity conditions P1-2, the population intervention indirect effect is given by, 
 \begin{align*}
	PIIE(a^*) = E[Y] - E[Y(Z(a^*))] = E[Y] - \Psi 
	\end{align*} 
	where \vspace{-.5cm}
	\begin{align}
	\Psi & = \sum_{z,c} Pr(Z = z \mid A=a^*, C=c) \notag \\
	& \hspace{2.5cm} \times  \sum_{a} E(Y \mid A=a, Z=z, C=c) Pr(A=a \mid C=c) Pr(C=c) \label{psi}
	\end{align}
\end{lemma}
\noindent Further, equation (\ref{psi}) implies nonparametric identification in the sense that conditions M1-3 and P1-2 do not restrict the observed data distribution. The proof for this lemma can be found in the Appendix section A1.1. 

Interestingly, $\Psi$ is closely connected to Judea Pearl's front-door criterion. Pearl's front-door criterion provides conditions for identification of the indirect effect in the presence of unmeasured confounding of the exposure-outcome relation. The criterion requires: (1) $Z$ intercepts all directed paths from the exposure $A$ to the outcome $Y$ so that the indirect effect equals the total effect of $A$ on $Y$, (2) there is no unblocked back-door path from $A$ to $Z$, and (3) all back-door paths from $Z$ to $Y$ are blocked by $A$ and $C$ \citep{pearl2009causality}. More formally, suppose that M1-3 and the following additional assumption hold, 
\begin{align*} \textrm{F1.} & \ Y(a,z) = Y(a^*,z) = Y(z) \ \ \  \forall \ a, a^*,z \end{align*}
F1 crucially states that $Z$ fully mediates the effect of $A$ on $Y$. In other words, mediator variable(s) $Z$ intercepts all directed paths from the exposure to the outcome. Figure 1b encodes one possible graph that satisfies the front-door criterion under a Finest Fully Randomized Causally Interpretable Structured Tree Graph, a submodel of the NPSEM-IE, interpretation of the causal diagram \citep{robins1986new, pearl2009causality, pearl2012causal}. 

When F1 holds, the term $E(Y(Z(a^*)))$ reduces to $E(Y(a^*))$. The identifying formula for the latter term is known as Pearl's front-door functional and matches equation (\ref{psi}) \citep{pearl2009causality}. See Appendix A2.1 for proof and further discussion. Under the front-door criterion (e.g. M1-3 and F1), the population intervention indirect effect can be expressed as, 
\begin{align}
PIIE(a^*) = E[Y] - E[Y(a^*)] = PIE(a^*) \label{frontdoor}
\end{align}
\noindent That is, the $PIIE(a^*)$ is equal to the $PIE(a^*)$ when F1 holds. The identifying conditions for the PIIE can be thought of as a generalization of Pearl's front-door criterion as F1 need not hold, thereby allowing a direct effect of the exposure $A$ on the outcome $Y$, not through the mediator variable(s) $Z$ (i.e. the PIDE may or may not be null). Importantly, while the PIIE is nonparametrically identified under  M1-3, the PIE and the PIDE are not identified. In the event that M4 also holds, and thus $E[Y(a^*)]$ is identified, the PIE and PIDE are both nonparametrically identified along with the NIE and PIIE. 

In the special case of binary $A$, the PIE can be written as the effect of treatment on the treated (ETT) scaled by prevalence of treated persons, 
\begin{align*}
PIE(0)& = \underbrace{E(Y(1)-Y(0)| A=1))}_\text{ETT} \times Pr(A=1) 
\end{align*}
See proof in Appendix section A2.5. Thus, the PIIE and PIDE can respectively be written as the indirect and direct components of the ETT simply upon rescaling by the prevalence of treated persons. This decomposition of the ETT offers an alternative to that of \cite{vansteelandt2012natural}. Further, in the case of binary $Y$, the PIE can be written as the attributable fraction (AF) scaled by the prevalence of outcome,
\begin{align*}
PIE(a^*)& =  \underbrace{ [  E(Y-Y(a^*))/E(Y) ] }_\text{AF}  \times E(Y) 
\end{align*}
Thus, the PIIE and PIDE can also be written as the indirect and direct components on the AF simply upon rescaling by prevalence of outcome. This decomposition of the AF offers an alternative to that of \cite{sjolander7mediation}. Further discussion can be found in Appendix section A2.6.

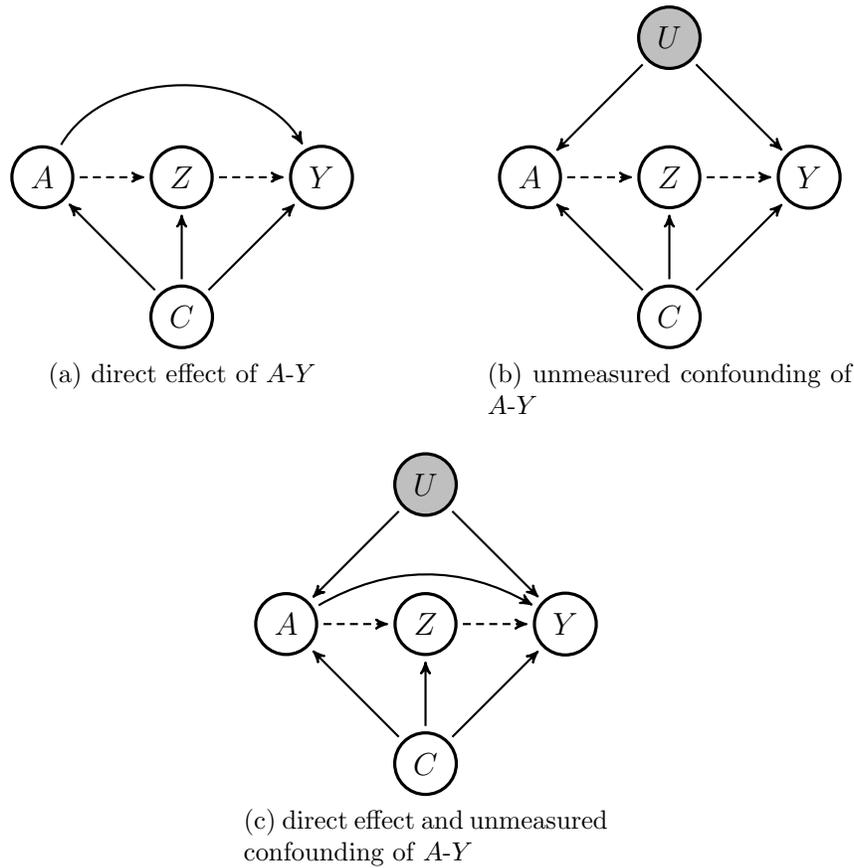
\begin{figure}[htbp!]
	\centering
	\subfloat[direct effect of $A$-$Y$]{
		\begin{tikzpicture}[->,>=stealth',node distance=1cm, auto,]
		\node[est] (A) {$A$};
		\node[est, right = of A] (Z) {$Z$};
		\node[est, right = of Z] (Y) {$Y$};
		\node[est, below = of Z] (C) {$C$};
		\path[pil, densely dashed] (A) edgenode {} (Z);
		\path[pil, densely dashed] (Z) edgenode {} (Y);
		\path[pil] (C) edgenode {} (A);
		\path[pil] (C) edgenode {} (Z);
		\path[pil] (C) edgenode {} (Y);
		\path[pil] (A) edge [bend left=60] node [left]  {} (Y);
		\end{tikzpicture} } \hspace{1.5cm}
	\subfloat[unmeasured confounding of $A$-$Y$]{
		\begin{tikzpicture}[->,>=stealth',node distance=1cm, auto,]
		\node[est] (A) {$A$};
		\node[est, right = of A] (Z) {$Z$};
		\node[est, right = of Z] (Y) {$Y$};
		\node[est, below = of Z] (C) {$C$};
		\node[shade, above = of Z] (U) {$U$};
		\path[pil,densely dashed]  (A) edgenode {} (Z);
		\path[pil,densely dashed]  (Z) edgenode {} (Y);
		\path[pil] (U) edgenode {} (A);
		\path[pil] (U) edgenode {} (Y);
		\path[pil] (C) edgenode {} (A);
		\path[pil] (C) edgenode {} (Z);
		\path[pil] (C) edgenode {} (Y);
		\end{tikzpicture} } \hspace{1.5cm}
	\subfloat[direct effect and unmeasured confounding of $A$-$Y$]{
		\begin{tikzpicture}[->,>=stealth',node distance=1cm, auto,]
		\node[est] (A) {$A$};
		\node[est, right = of A] (Z) {$Z$};
		\node[est, right = of Z] (Y) {$Y$};
		\node[est, below = of Z] (C) {$C$};
		\node[shade, above = of Z] (U) {$U$};
		\path[pil,densely dashed]  (A) edgenode {} (Z);
		\path[pil,densely dashed]  (Z) edgenode {} (Y);
		\path[pil] (U) edgenode {} (A);
		\path[pil] (U) edgenode {} (Y);
		\path[pil] (C) edgenode {} (A);
		\path[pil] (C) edgenode {} (Z);
		\path[pil] (C) edgenode {} (Y);
		\path[pil] (A) edge [bend left] node  {} (Y);
		\end{tikzpicture} }
	\caption{Causal diagrams with indirect effects as dashed lines. The following indirect effects are identified in each diagram under a Nonparametric Structural Equation Model with Independent Errors \citep{pearl2009causality} interpretation of the diagram: (a)  natural indirect effect and population intervention indirect effect, (b) natural indirect effect (equal to the total effect) and population intervention indirect effect (equal to the population intervention effect), and (c) population intervention indirect effect. Further, the indirect effects in (b) are identified under a Finest Fully Randomized Causally Interpretable Structured Tree Graph \citep{robins1986new}, which does not encode so-called ``cross-world" assumptions such as M3.} \label{figure1}
\end{figure}

\section{Estimation and Inference}

\subsection{Parametric estimation}

We have considered identification under a nonparametric model for the observed data distribution. Estimation of formula (\ref{psi}) clearly requires estimation of the mean of $Y|A, Z, C$ and the densities for $Z|A,C$, $A|C$, and $C$. In principle, one may wish to estimate these quantities nonparameterically; however, as will typically be the case in practice, the observed set of covariates $C$ may have two or more components that are continuous, so that the curse of dimensionality would rule out the use of nonparametric estimators such as kernel smoothing or series estimation. Thus, we propose four estimators for the population intervention indirect effect that impose parametric models for different parts of the observed data likelihood, allowing other parts to remain unrestricted. Under this setting, each estimator will be consistent and asymptotically normal (CAN) under the assumed semiparametric model. We also propose a doubly robust estimator which is CAN under a semiparametric union model thereby allowing for robustness to partial model misspecification. 

We only discuss estimation for the second term in the PIIE contrast, $\Psi$, as the first term $E(Y)$ can be consistently estimated nonparametrically by the empirical mean of $Y$. Let $Pr(y|a,z,c; \theta)$ denote a model for the density of $Y|A, Z, C$ evaluated at $y, a, z, c$ and indexed by $\theta$. Likewise, let $Pr(z|a,c; \beta)$and $Pr(a|c; \alpha)$ denote models for $Z|A,C$ and $A|C$ evaluated at $z,a,c$ and $a,c$ respectively with corresponding parameters $\beta$ and $\alpha$. These models could in principle be made as flexible as allowed by sample size, to simplify exposition, we will focus on simple parametric models. The first of the four estimators is the maximum likelihood estimator (MLE), $\hat{\Psi}_{mle}$, under a model that specifies parametric models for $A$, $Z$, and $Y$, and a nonparametric model for the distribution of $C$ estimated by its empirical distribution. The MLE is obtained by the plug-in principle \citep{casella2002statistical}: 
\begin{align*}
\hat{\Psi}_{mle} & =\frac{1}{n} \sum_{i=1}^n \bigg\{ \sum_{z} Pr(Z = z \mid A=a^*, C_i ; \hat{\beta}) \times \\
 & \hspace{3cm}  \sum_{a} E(Y \mid A=a, Z=z, C_i ; \hat{\theta}) Pr(A=a \mid C_i ; \hat{\alpha}) \bigg\}
\end{align*} 
\noindent where $\hat{\theta}$, $\hat{\beta}$, and $\hat{\alpha}$ are the MLEs of $\theta$, $\beta$, and $\alpha$. This estimator is only consistent under correct specification of the three required models, which we define as $\mathcal{M}_{y,z,a}$. For the remainder of the paper, we consider an alternate ML estimator under model $\mathcal{M}_{y,z}$, which specifies parametric models for $Z$ and $Y$, and a nonparametric model for the joint distribution of $A,C$ estimated by its empirical distribution. 
\begin{align*}
\hat{\Psi}^{alt}_{mle} & =\frac{1}{n} \sum_{i=1}^n \bigg\{ \sum_{z} Pr(Z = z \mid A=a^*, C_i ; \hat{\beta}) E(Y \mid A_i, Z=z, C_i ; \hat{\theta}) \bigg\}
\end{align*}

\subsection{Semiparametric estimation} 

Next, we consider two semiparametric estimators for $\Psi$. The first is under model $\mathcal{M}_z$ which posits a density for the law of $Z|A,C$ but allows the densities of $Y|A,Z,C$, $A | C$, and $C$ to remain unrestricted. The second is under model $\mathcal{M}_{y,a}$ which instead posits a density for the outcome mean of $Y|Z, A, C$ and the density of $A|C$ but allows the densities of $Z|A,C$ and $C$ to be unrestricted. 
\begin{align*}
\hat{\Psi}_1 & = \frac{1}{n} \sum_{i=1}^n Y_i \frac{f(Z_i \mid a^*, C_i; \hat{\beta})}{f(Z_i \mid A_i, C_i; \hat{\beta})} \\
\hat{\Psi}_2 & = \frac{1}{n} \sum_{i=1}^n \frac{I(A_i = a^*)}{f(A_i \mid C_i; \hat{\alpha})} E\big( E \big\{ Y_i \mid A_i,Z_i,C_i ; \hat{\theta} \big\} \mid C_i; \hat{\alpha} \big) 
\end{align*}
\begin{lemma} Under standard regularity conditions and P1, the estimator $\hat{\Psi}_1$ is consistent and asymptotically normal under model $\mathcal{M}_{z}$. \end{lemma}
\begin{lemma} Under standard regularity conditions and P2, the estimator $\hat{\Psi}_2$ is consistent and asymptotically normal under model $\mathcal{M}_{y,a}$. \end{lemma}

The estimator $\hat{\Psi}_1$ will generally fail to be consistent if the density for $Z | A,C$ is incorrectly specified even if the rest of the likelihood is correctly specified. Likewise, the estimator $\hat{\Psi}_2$ will also generally fail to be consistent if either the mean model for $Y|A,Z,C$ or the density of $A|C$ is incorrectly specified. In order to motivate our doubly robust estimator, the following result gives the efficient influence function for $\Psi$ in the nonparametric model $\mathcal{M}_{np}$, which does not place any model restriction on the observed data distribution. The following results are entirely novel and have previously not appeared in the literature. 

\setcounter{theorem}{0}
\begin{theorem}
	The efficient influence function of $\Psi$ in $\mathcal{M}_{np}$ is:
	\begin{align} 
	\varphi^{eff}(Y, Z, A, C) & = (Y - E(Y \mid A,Z,C)) \frac{f(Z \mid a^*, C)}{f(Z \mid A,C)} \notag \\
	& \hspace{.5cm} + \frac{I(A = a^*)}{f(A \mid C)} \big( \sum_a E[Y \mid a, Z, C] f(a \mid C) \notag \\
	& \hspace{4cm} - \sum_{a, \bar{z}} E(Y \mid a, \bar{z}, C) f(\bar{z} \mid A, C) f(a \mid C) \big) \notag \\
	&\hspace{.5cm}  + \sum_{z} E[Y \mid A, z, C] f(z \mid a^*,C)   - \Psi \label{eif}
 \end{align}
	\noindent and the semiparametric efficiency bound of $\Psi$ in $\mathcal{M}_{np}$ is given by $var\{ \varphi^{eff} \}$. 
\end{theorem}
The proof for this theorem can be found in the Appendix section A1.4. An implication of this result is that for any regular and asymptotically linear (RAL) estimator $\hat{\Psi}$ in model $\mathcal{M}_{np}$ it must be that $\sqrt{n}( \hat{\Psi} - \Psi) = \frac{1}{\sqrt{n}} \sum_{i=1}^n \varphi^{eff}(Y_i, Z_i, A_i, C_i) + o_p(1)$. In other words, all RAL estimators in this model are asymptotically equivalent and attain the semiparametric efficiency bound of $\Psi$ in $\mathcal{M}_{np}$ \citep{bickel1998efficient}.  The result motivates the following estimator of $\Psi$, which we formally establish to be doubly robust. 

\begin{align}
\hat{\Psi}_{dr} & = \frac{1}{n} \sum_{i=1}^n  [Y_i  - E(Y \mid A_i,Z_i,C_i; \hat{\theta})] \frac{f(Z \mid a^*, C_i; \hat{\beta})}{f(Z \mid A_i,C_i; \hat{\beta})} \notag \\
& \hspace{1cm} + \frac{I(A_i = a^*)}{Pr(A_i =a^* \mid C_i; \hat{\alpha})} \big( \sum_a E[Y \mid a, Z_i, C_i; \hat{\theta}] f(a \mid C_i; \hat{\alpha}) \notag  \\
& \hspace{7cm} - \sum_{a, \bar{z}} E(Y \mid a, \bar{z}, C_i; \hat{\theta}) f(\bar{z} \mid A_i, C_i; \hat{\beta}) f(a \mid C_i; \hat{\alpha}) \big)  \notag \\
& \hspace{2cm} + \sum_{z} E[Y \mid A_i, z, C_i; \hat{\theta}] f(z \mid a^*,C_i; \hat{\beta})  \label{est_sp}
\end{align}
\begin{theorem}
	Under standard regularity conditions and the positivity assumptions given by P1 and P2, the estimator $\hat{\Psi}_{dr}$ is consistent and asymptotically normal provided that one of the following holds: (1) the model for the mean $E(Y | A,C)$ and the exposure density $f(A | C)$ are both correctly specified; or (2) The model for the mediator density $f(Z | A,C)$ is correctly specified. Also, $\hat{\Psi}_{dr}$ attains the semiparametric efficiency bound for the union model $\mathcal{M}_{union} = \mathcal{M}_{y,a} \bigcup \mathcal{M}_{z}$, and therefore for the nonparametric model $\mathcal{M}_{np}$ at the intersection submodel where all models are correctly specified. 
\end{theorem}

The estimator $\hat{\Psi}_{dr}$ offers two genuine opportunities to consistently estimate $\Psi$, and, thus, the PIIE. This is clearly an improvement over the other estimators $\hat{\Psi}_{mle}$, $\hat{\Psi}_1$ and $\hat{\Psi}_2$, which are only guaranteed to be consistent under more stringent parametric restrictions. In addition, the doubly-robust estimator achieves the semiparametric efficiency bound in the union model $\mathcal{M}_{union}$ and will thus have valid inference provided one of the two strategies holds. Note that the estimator will be less efficient than the MLE in the submodel $\mathcal{M}_{y,z,a}$ where all models are correctly specified. For inference on $\Psi$, we provide a consistent estimator of the asymptotic variance for the proposed estimators in the Appendix section A2.4. Wald-type confidence intervals for $\Psi$ can then be based on $\hat{\Psi}_{mle}$, $\hat{\Psi}_1$, $\hat{\Psi}_2$, or $\hat{\Psi}_{dr}$ and the corresponding standard error estimator. 

An important advantage of the doubly-robust estimator is that it can easily accommodate modern machine learning for estimation of high dimensional nuisance parameters, such as $E(Y|A,Z,C)$ or $f(Z|A,C)$ \citep{van2011targeted,newey2017cross,chernozhukov2017double}. Although, investigators should exercise caution when implementing these more flexible methods, particularly if nonparametric methods are used to estimate nuisance parameters. This is because such methods typically cannot attain root-$n$ convergence rates, although the doubly robust estimator would in principle provide valid root-n inferences about $\Psi$ provided that estimators of nuisance parameters have a convergence rate faster than $n^{-1/4}$ \citep{newey1990semiparametric,robins2017higher}. A major challenge with using complex machine learning methods such as random forests arises if the corresponding estimator of the nuisance function (say $f(A|C)$) fails to be consistent at rate $n^{1/4}$ even if the other nuisance function (say $f(Z|A,C)$) is estimated at rate root-$n$, in such case, it is not entirely clear what the asymptotic distribution is for $\hat{\Psi}_{dr}$.

\section{Simulation Study}

\subsection{Data generating mechanism}

We now report extensive simulation studies which aim to illustrate: (i) robustness of PIIE to exposure-outcome unmeasured confounding (ii) robustness properties to model misspecification of our various semiparametric estimators. The data generating mechanism for simulations was as followed: 
\begin{align*}
C_1 & \sim Ber(.6) \\
C_2 | C_1 & \sim Ber(\textrm{expit}(1 + .5c_1)) \\
C_3 & \sim Ber(.3) \\
A | C_1, C_2, C_3 & \sim Ber(\textrm{expit}(.5 + .2c_1 + .4c_2 + .5c_1 c_2 + .2 c_3)) \\
Z | A, C_1, C_2 & \sim N(1 + a -2c_1 + 2c_2 + 8c_1 c_2, \ 4) \\
Y | A, Z, C_1, C_2, C_3 & \sim N(1 + 2a + 2z - 8az + 3c_1 + c_2 + c_1 c_2 + c_3, \ 1)
\end{align*}
Therefore, $C_1$, $C_2$, and $C_3$ confound the $A-Y$ association while only $C_1$ and $C_2$ confound the $A-M$ and $M-Y$ associations. Simulations were performed 10,000 times with a sample size of 1,000. We evaluated the performance of the proposed estimators under the following settings,
\begin{align*}
(a) &\ \mathcal{M}_{y,z,a} : \ \stackrel{*}{E}(Y \mid a, z, c_1, c_2, c_3), \ \stackrel{*}{f}(Z \mid a,c_1,c_2), \ \stackrel{*}{f}(A \mid c_1,c_2,c_3) \\
(b) &\ \mathcal{M}_{y,z,a}': \overline{E}(Y \mid a, z, c_1, c_2) \ \textrm{($c_3$ left out)}, \ \overline{f}(A \mid c_1,c_2) \ \textrm{($c_3$ left out)},  \ \stackrel{*}{f}(Z \mid a,c_1,c_2) \\
(c) &\ \mathcal{M}_{z}: \ \ \ \tilde{E}(Y \mid a, z, c_1, c_2) \ \textrm{($az$ left out)}, \ \tilde{f}(A \mid c_1) \ \textrm{($c_2$, $c_1c_2$, $c_3$ left out)}, \ \stackrel{*}{f}(Z \mid a,c_1,c_2) \\
(d) &\ \mathcal{M}_{y,a}: \ \ \stackrel{*}{E}(Y \mid a, z, c_1, c_2), \ \stackrel{*}{f}(A \mid c_1), \ \tilde{f}(Z \mid a,c_1) \ \textrm{($c_2$, $c_1c_2$, $c_3$ left out)}
\end{align*}
\noindent where $*$ denotes that the model is correctly specified and $\sim$ and $-$ denote the model is misspecified. Note that the alternate ML estimator,  $\hat{\Psi}^{alt}_{mle}$, does not specify a model for $A \mid C$. 

\subsection{Results}
Estimation and inference were performed using the \texttt{piieffect} function implemented in the \texttt{frontdoorpiie} R package \citep{fulchergithub}. Under simple linear models for the outcome and mediator variables, the variance estimator of the MLE admits a simple closed form expression (see Appendix section A2.3). The variance estimator for the semiparametric estimators is described in Appendix section A2.4. Alternatively, one may use the nonparametric bootstrap for inference. 

In both Figure 2 and Table 1, the maximum likelihood estimator $\hat{\Psi}^{alt}_{mle}$ was only consistent under correct model specification (a) whether or not there was unmeasured confounding of the exposure-outcome relationship (b). This confirms our theoretical result as the PIIE is in fact empirically identified even if the exposure-outcome relationship is subject to unmeasured confounding. The MLE is not robust to model misspecification of the form in scenarios (c) and (d). On the other hand, the doubly-robust semiparametric estimator $\hat{\Psi}_{dr}$ appears to be consistent under all scenarios (a)-(d). The semiparametric estimator $\hat{\Psi}_{1}$ which only depends on the choice of model for the density for $Z | A,C$ has large bias in scenario (d). The semiparametric estimator $\hat{\Psi}_{2}$ which only depends on a model for the mean $Y | A,Z,C$ and $A | C$ has large bias in scenario (c). As expected, the maximum likelihood estimator is more efficient than the semiparametric estimators when all parametric models are correctly specified. For correctly specified models, Monte Carlo coverage of 95\% confidence intervals was close to nominal level. Confidence intervals based on inconsistent estimators had incorrect coverage. 

\setcounter{figure}{1}
\begin{figure}[!htbp]
	\centering
	\caption{Population intervention indirect effect by estimator and model specifications} 	\includegraphics[scale=.11] {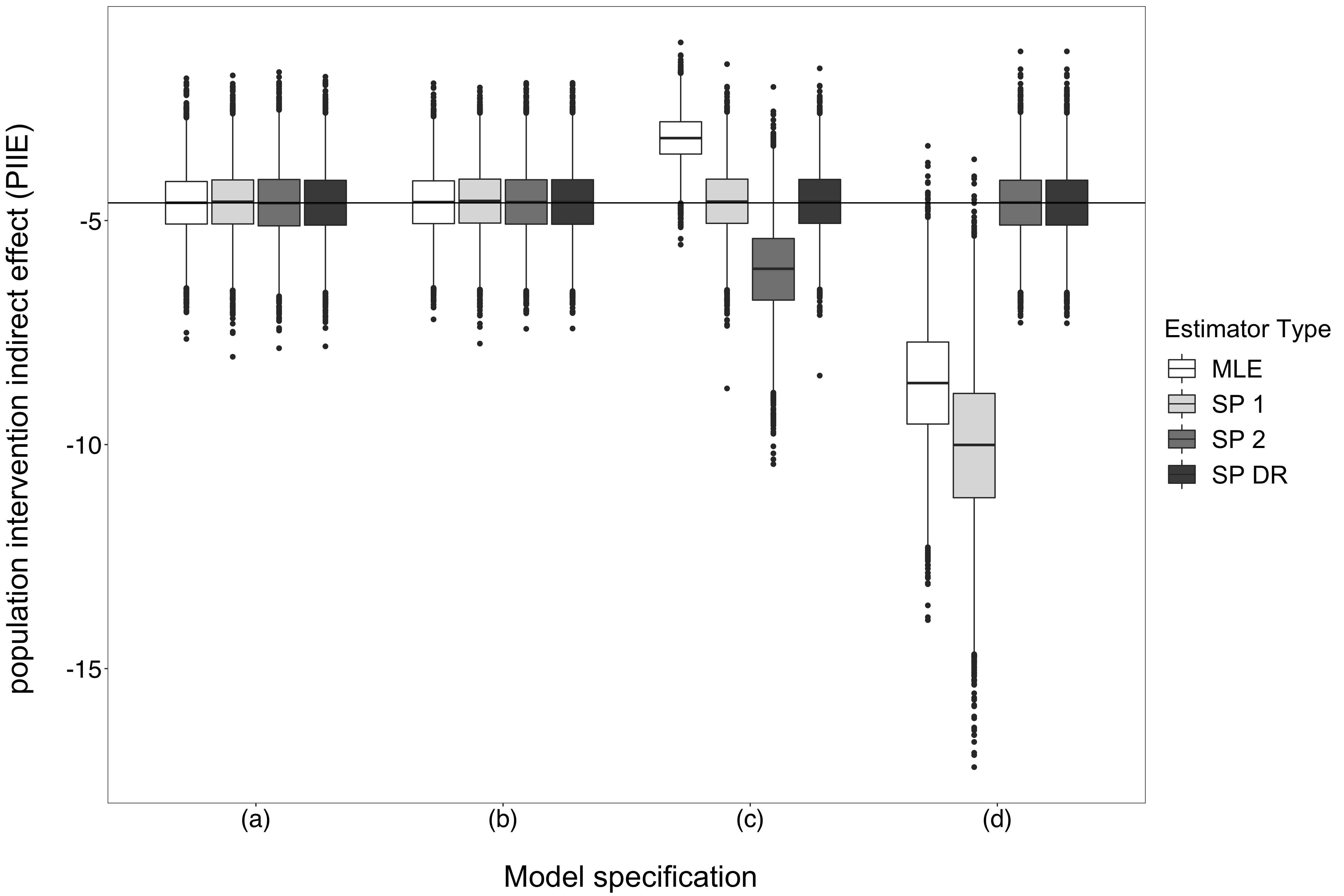}
\end{figure}

\begin{table}[!htbp] \centering 
	 \caption{Operating characteristics by model specifications and estimator} 
	 \label{} 
	 \fbox{%
	\begin{tabular}{lccccc}
		& $\hat{\Psi}$ & $\widehat{PIIE}$ & Variance & Proportion bias & .95 CI Coverage \\ 
		\hline
		&&&&& \\
	    \large $\mathcal{M}_{y,z,a}$ \normalsize &&&&& \\
		MLE & -18.19 & -4.61 & 0.50 & $<0.01$ & 0.95 \\ 
		SP 1 & -18.20 & -4.59 & 0.54 & $<0.01$ & 0.95 \\ 
		SP 2 & -18.19 & -4.61 & 0.60 & $<0.01$ & 0.95 \\ 
		SP DR & -18.19 & -4.61 & 0.56 & $<0.01$ & 0.95 \\ 
		&&&&& \\
		\large $\mathcal{M}_{y,z,a}'$\normalsize  &&&&& \\ 
		MLE  & -18.20 & -4.59 & 0.50 & $<0.01$& 0.95 \\ 
		SP 1 & -18.21 & -4.57 & 0.54 & $<0.01$ & 0.95 \\ 
		SP 2 & -18.20 & -4.59 & 0.55 & $<0.01$ & 0.94 \\ 
		SP DR & -18.20 & -4.59 & 0.55 & $<0.01$ & 0.94 \\ 
		&&&&& \\
		\large $\mathcal{M}_{z}$\normalsize  &&&&& \\ 
		MLE & -19.63 & -3.17 & 0.27 & -0.31 & 0.23 \\ 
		SP 1 & -18.22 & -4.58 & 0.54 & $<0.01$ & 0.95 \\ 
		SP 2 & -16.70 & -6.10 & 1.04 & 0.33 & 0.70 \\ 
		SP DR & -18.22 & -4.58 & 0.52 & $<0.01$ & 0.94 \\
		&&&&& \\
		\large $\mathcal{M}_{y,a}$ \normalsize &&&&& \\
		MLE & -14.16 & -8.64 & 1.61 & 0.88 & 0.12 \\ 
		SP 1 & -12.75 & -10.05 & 3.32 & 1.18 & 0.10 \\ 
		SP 2 & -18.20 & -4.60 & 0.55 & $<0.01$ & 0.95 \\ 
		SP DR & -18.20 & -4.60 & 0.55 & $<0.01$ & 0.95 \\
	\end{tabular}}
			\footnotesize 
			\begin{flushleft}
			Note: for the $\hat{\Psi}$ column, MLE refers to using the $\hat{\Psi}^{alt}_{mle}$ estimator for the $\widehat{PIIE}$. Likewise, SP1 refers to using $\hat{\Psi}_{1}$, SP2 refers to using $\hat{\Psi}_{2}$, and SP DR refers to using $\hat{\Psi}_{dr}$ \\
			\end{flushleft}
\end{table}
\newpage
\section{Safer Deliveries Program in Zanzibar, Tanzania}
The Safer Deliveries program aimed to reduce the high rates of maternal and neonatal mortality in Zanzibar, Tanzania by increasing the number of pregnant women who deliver in a health care facility and attend prenatal and postnatal check-ups. As of May 2017, the program was active in six (out of 11) districts in Zanzibar on the islands of Unguja and Pemba. The program trains community health workers (CHWs) selected by the Ministry of Health to participate in the program based on their literacy, expressed commitment to the improvement of health, and respectability in their communities. 

The CHWs work with community leaders and staff at nearby health facilities to identify and register pregnant women and are expected to visit the woman in her home three times during pregnancy to screen for danger signs and provide counseling to help the woman prepare for a facility delivery. During the registration visit, the mobile app calculated a woman's risk category (low, medium, or high) based on a combination of obstetric and demographic factors. Women categorized as high risk were instructed to deliver at a referral hospital. The app then calculated a recommended savings amount based on the women's recommended delivery location. On average, high risk women were recommended to save more money than low or medium risk women as they were recommended to deliver at referral hospitals of which there are only four on the island. This analysis assessed the effectiveness of this tailored savings recommendation by risk category on actual savings.

We considered high risk category (vs. low or medium risk) as our binary exposure of interest; although, our methods would equally apply for categorical exposure variable. The mediator variable was recommended savings in Tanzanian Shilling (TZS), which was calculated during the first visit. The outcome variable was actual savings achieved by the woman and her family at time of her delivery. In the analysis, we adjusted for district of residence to account for regional differences in health-seeking behavior and accessibility of health facilities. The population intervention indirect effect was the best estimand for this research question as we were interested in the mediated effect of savings recommendations under the risk categories observed in the current population. Additionally, there was likely unmeasured confounding between the exposure (high risk) and outcome (actual savings) relationship because most socio-economic factors and health-seeking behavior that may be associated with other factors related to risk category and a woman's ability to save were not collected by the program. Furthermore, confounding of exposure-mediator and mediator-outcome associations was less of a concern as the app calculated the recommended savings based on the delivery location which is determined both by risk category and distance to the appropriate health facility. That is, women who are in a low risk category are recommended to deliver at the facility closest to them, whereas women in the high risk category are recommended to deliver at one of four available referral facilities in Zanzibar. 

\begin{table}[!htbp] \centering 
	\caption{Characteristics of the Safer Deliveries study population (n=4,102)} 
	\label{} 
	\fbox{%
		\begin{tabular}{lr}
			\textbf{Variable} &  $n$ (\%) \\ 
			\hline
			\textbf{Risk category}  & \\
			\ \ Low or medium & 3,364 (82)  \\ 
			\ \ High & 738 (18)  \\
			\textbf{District}   &  \\ 
			\ \ North A & 977 (24)  \\ 
			\ \ North B & 1,392 (34)  \\ 
			\ \ Central & 691 (17)  \\ 
			\ \ West & 798 (19)  \\ 
			\ \ South & 244 (6)  \\ 
			\textbf{Recommended savings} & \\ 
			\ \ mean (sd) & 13.12 (6.03) \\
			\textbf{Actual savings} &  \\ 
			\ \ mean (sd) & 14.09 (12.11)
	\end{tabular}}
\end{table}

This study included women enrolled in the Safer Deliveries program who had a live birth by May 31, 2017 (n=4,511). We excluded: 253 women from the newly-added Mkoani district of Pemba Island, 2 women with missing LMP date and EDDs, 31 women with invalid enrollment times, and 123 women with missing risk category, district, or savings information. Our final study population included 4,102 women. Therefore, the following analyses are only valid under an assumption that data are missing completely at random. The observed average savings at time of delivery was \$14.09. Note that for ease of interpretation we converted from Tanzanian Shilling (1 USD = 2,236.60 TZS on May 31, 2017). We estimated the population intervention indirect effect; that is, the difference in average savings between the current population of women and a population of women had possibly contrary to the fact every woman received the savings recommendation of a low or medium risk woman. To estimate the population intervention indirect effect we employed our four estimators under the following parametric models:
\begin{align*}
& highrisk = \alpha_0 + \alpha_2^T district + \varepsilon_a \\
& savings_{rec} = \beta_0 + \beta_1 highrisk + \beta_2^T district + \varepsilon_z \\
& savings_{act}  = \theta_0 + \theta_1 highrisk  + \theta_2 savings_{rec} + \theta_3^T district + \varepsilon_y 
\end{align*}
Table 2 gives the distribution of variables in this study population. The maximum likelihood estimator, $\hat{\Psi}^{alt}_{mle}$, estimated the average savings for all women had their recommended savings been set to the amount they would have been recommended to save had they not been high risk to be \$13.87 resulting in a PIIE of \$0.22 with a 95\% CI of (\$0.15, \$0.30). The semiparametric estimator that only includes models for $A | C$ and $Y | A, Z, C$, $\hat{\Psi}_{2}$, gave almost identical results. The doubly robust semiparametric estimator of the PIIE was estimated for to be \$13.95 with 95\% CI of (-\$0.03, \$0.32). The semiparametric estimator that only depends on a parametric model for $Z | A, C$, $\hat{\Psi}_{1}$ resulted in very similar inferences to the doubly-robust estimator. To compare these estimators, we conducted a bootstrap test of the null hypothesis that each of the estimator (MLE, SP1, SP2) converged to the same probability limit as the semiparametric doubly-robust estimator. The procedure was motivated by \cite{hausman1978specification} to directly test whether two estimators are consistently estimating the same parameter value. We used 1,000 boostrap samples and did not find evidence of a difference between any of the three estimators and the SP DR ($P=0.35$ for MLE; $P=0.14$ for SP1; $P=0.36$ for SP2). As such, we concluded that there was evidence of a non-zero PIIE -- revealing that the tailored savings recommendations to high risk women affects their actual savings by the time of their delivery. On average, if high risk women had been recommended to save what they would have if they were low to medium risk, this would slightly decrease the amount of money she saved. 

\begin{table}[!htbp]  
	\caption{Effect of risk category on actual savings mediated by recommended savings ($n=4,102$) }
	\centering
	\fbox{%
	\begin{tabular}{ccccc} 
		& $\hat{\Psi}$ & $\widehat{PIIE}$ & Standard Error & 95\% CI \\ 
		\hline
		MLE  & 13.87 & 0.22 & 0.04 & (0.15, 0.30) \\ 
		SP 1   & 14.08 & 0.02 & 0.11 & (-0.20, 0.23) \\ 
		SP 2  & 13.87 & 0.22 & 0.05 &  (0.13, 0.31)  \\ 
		SP DR & 13.95 & 0.14 & 0.09 & (-0.03, 0.32) 
	\end{tabular}}
\end{table}
\newpage
\section{Discussion}

In this paper, we have presented a decomposition of the population intervention effect, which we have argued is useful to address policy-related questions at the population-level especially in the presence of a harmful exposure. In addition, the decomposition offers an alternative to the recently proposed decompositions for the effect of treatment on the treated \citep{vansteelandt2012natural} and the attributable fraction \citep{sjolander7mediation}. Importantly, our resulting population intervention indirect effect is robust to unmeasured confounding of the exposure-outcome relationship, which does not hold for the natural indirect effect, natural indirect effect on the exposed, nor the natural indirect attributable fraction. We note that in a separate manuscript, we recently established that the NIE can in fact be identified if one replaces M4 with the assumption that there is no additive interaction between the mediator and the unmeasured confounder of the $A-Y$ association, a strictly stronger requirement than that for the PIIE \citep{fulcher2018estimation}.

We developed a doubly-robust estimator for the PIIE, which is consistent and asymptotically normal in a union model where at least one of the following hold: (1) outcome and exposure models are correctly specified or (2) mediator model is correctly specified. Our estimator is strictly more robust than the multiply robust estimator for the NIE proposed by \cite{tchetgen2012semiparametric}, which requires that any two of the three models is correctly specified. \cite{sjolander7mediation} proposed a doubly-robust estimator for the natural indirect attributable fraction requiring that either $p(Y |A,M,C)$ or $p(A|M,C)$ are correctly specified \underline{and} either $p(Y|A,C)$ or $p(A|C)$ are correctly specified. As mentioned by \cite{sjolander7mediation}, a doubly-robust estimator may not be realizable due to the fact various submodels of the union models are not variation independent, such that misspecification of the former generally rules out possibility that the latter could still be correctly specified. For example, when $M$ is binary, a logistic model for $p(Y|A,M,C)$ would imply a complex form for $p(Y | A,C)$. In a separate strand of work, \cite{lendle2013identification} developed an estimator for the natural indirect effect among the (un)exposed with the same robustness properties as \cite{sjolander7mediation}.

We emphasize that the use of the doubly-robust estimator of the PIIE does not obviate concerns about unmeasured confounding of the exposure-mediator, mediator-outcome relation, or exposure-induced mediator-outcome confounding. When such confounding is of concern, a sensitivity analysis should be performed \citep{vanderweele2011bias,tchetgen2012semiparametric,tchetgen2014estimation}. Investigators should exercise caution if they also wish to report the PIDE and PIE as these effects are not robust to exposure-outcome confounding. If exposure-outcome unmeasured confounding can be ruled out with reasonable certainty, then one can estimate the PIDE using our doubly-robust estimator for $\Psi$ and the well-known doubly-robust estimator for $E(Y(a^*))$ from \cite{robins2000sensitivity}. Likewise, the PIE can be estimated using the doubly-robust estimator developed by \cite{hubbard2008population}. 

Lastly, despite the front-door criterion being available in the literature for several years, this is the first methodology developed for semiparametric estimation and inference of the front-door functional $\Psi$. Therefore, when an investigator believes she has identified one or more mediator variables that satisfy the front-door criterion, she can use our proposed methodology to obtain an estimate of the PIE or the average causal effect that is not only doubly-robust, but also robust to unmeasured confounding of the exposure-outcome relation.

\bibliography{mybib}

\begin{thebibliography}{35}
\providecommand{\natexlab}[1]{#1}
\providecommand{\url}[1]{\texttt{#1}}
\expandafter\ifx\csname urlstyle\endcsname\relax
  \providecommand{\doi}[1]{doi: #1}\else
  \providecommand{\doi}{doi: \begingroup \urlstyle{rm}\Url}\fi

\bibitem[Angrist et~al.(1996)Angrist, Imbens, and
  Rubin]{angrist1996identification}
J.~D. Angrist, G.~W. Imbens, and D.~B. Rubin.
\newblock Identification of causal effects using instrumental variables.
\newblock \emph{Journal of the American statistical Association}, 91\penalty0
  (434):\penalty0 444--455, 1996.

\bibitem[Avin et~al.(2005)Avin, Shpitser, and Pearl]{avin2005identifiability}
C.~Avin, I.~Shpitser, and J.~Pearl.
\newblock Identifiability of path-specific effects.
\newblock \emph{Department of Statistics, UCLA}, 2005.

\bibitem[Bickel et~al.(1998)Bickel, Klaassen, Bickel, Ritov, Klaassen, Wellner,
  and Ritov]{bickel1998efficient}
P.~J. Bickel, C.~A. Klaassen, P.~J. Bickel, Y.~Ritov, J.~Klaassen, J.~A.
  Wellner, and Y.~Ritov.
\newblock \emph{Efficient and adaptive estimation for semiparametric models},
  volume~2.
\newblock Springer New York, 1998.

\bibitem[Campbell and Stanley(2015)]{campbell2015experimental}
D.~T. Campbell and J.~C. Stanley.
\newblock \emph{Experimental and quasi-experimental designs for research}.
\newblock Ravenio Books, 2015.

\bibitem[Casella and Berger(2002)]{casella2002statistical}
G.~Casella and R.~L. Berger.
\newblock \emph{Statistical inference}, volume~2.
\newblock Duxbury Pacific Grove, CA, 2002.

\bibitem[Chernozhukov et~al.(2017)Chernozhukov, Chetverikov, Demirer, Duflo,
  Hansen, Newey, and Robins]{chernozhukov2017double}
V.~Chernozhukov, D.~Chetverikov, M.~Demirer, E.~Duflo, C.~Hansen, W.~Newey, and
  J.~Robins.
\newblock Double/debiased machine learning for treatment and structural
  parameters.
\newblock \emph{The Econometrics Journal}, 2017.

\bibitem[Fulcher(2017)]{fulchergithub}
I.~Fulcher.
\newblock frontdoorpiie.
\newblock \url{https://github.com/isabelfulcher/frontdoorpiie}, 2017.

\bibitem[Fulcher et~al.(2018)Fulcher, Shi, and
  Tchetgen~Tchetgen]{fulcher2018estimation}
I.~R. Fulcher, X.~Shi, and E.~J. Tchetgen~Tchetgen.
\newblock Estimation of natural indirect effects robust to unmeasured
  confounding and mediator measurement error.
\newblock \emph{arXiv preprint arXiv:1808.03692}, 2018.

\bibitem[Geneletti and Dawid(2011)]{geneletti2007defining}
S.~Geneletti and A.~P. Dawid.
\newblock Defining and identifying the effect of treatment on the treated.
\newblock In P.~M.~I. Illari, F.~Russo, and J.~Williamson, editors,
  \emph{Causality in the Sciences}, chapter~34. Oxford Scholarship Online,
  2011.

\bibitem[Greenland and Drescher(1993)]{greenland1993maximum}
S.~Greenland and K.~Drescher.
\newblock Maximum likelihood estimation of the attributable fraction from
  logistic models.
\newblock \emph{Biometrics}, pages 865--872, 1993.

\bibitem[Hahn(1998)]{hahn1998role}
J.~Hahn.
\newblock On the role of the propensity score in efficient semiparametric
  estimation of average treatment effects.
\newblock \emph{Econometrica}, pages 315--331, 1998.

\bibitem[Hausman(1978)]{hausman1978specification}
J.~A. Hausman.
\newblock Specification tests in econometrics.
\newblock \emph{Econometrica: Journal of the econometric society}, pages
  1251--1271, 1978.

\bibitem[Hubbard and Van~der Laan(2008)]{hubbard2008population}
A.~E. Hubbard and M.~J. Van~der Laan.
\newblock Population intervention models in causal inference.
\newblock \emph{Biometrika}, 95\penalty0 (1):\penalty0 35--47, 2008.

\bibitem[Imai et~al.(2010{\natexlab{a}})Imai, Keele, and
  Tingley]{imai2010general}
K.~Imai, L.~Keele, and D.~Tingley.
\newblock A general approach to causal mediation analysis.
\newblock \emph{Psychological methods}, 15\penalty0 (4):\penalty0 309,
  2010{\natexlab{a}}.

\bibitem[Imai et~al.(2010{\natexlab{b}})Imai, Keele, and
  Yamamoto]{imai2010identification}
K.~Imai, L.~Keele, and T.~Yamamoto.
\newblock Identification, inference and sensitivity analysis for causal
  mediation effects.
\newblock \emph{Statistical science}, pages 51--71, 2010{\natexlab{b}}.

\bibitem[Imbens and Lemieux(2008)]{imbens2008regression}
G.~W. Imbens and T.~Lemieux.
\newblock Regression discontinuity designs: A guide to practice.
\newblock \emph{Journal of econometrics}, 142\penalty0 (2):\penalty0 615--635,
  2008.

\bibitem[Lendle et~al.(2013)Lendle, Subbaraman, and van~der
  Laan]{lendle2013identification}
S.~D. Lendle, M.~S. Subbaraman, and M.~J. van~der Laan.
\newblock Identification and efficient estimation of the natural direct effect
  among the untreated.
\newblock \emph{Biometrics}, 69\penalty0 (2):\penalty0 310--317, 2013.

\bibitem[Lipsitch et~al.(2010)Lipsitch, Tchetgen~Tchetgen, and
  Cohen]{lipsitch2010negative}
M.~Lipsitch, E.~Tchetgen~Tchetgen, and T.~Cohen.
\newblock Negative controls: a tool for detecting confounding and bias in
  observational studies.
\newblock \emph{Epidemiology (Cambridge, Mass.)}, 21\penalty0 (3):\penalty0
  383, 2010.

\bibitem[Miao and Tchetgen~Tchetgen(2017)]{miao2017invited}
W.~Miao and E.~Tchetgen~Tchetgen.
\newblock Invited commentary: bias attenuation and identification of causal
  effects with multiple negative controls.
\newblock \emph{American journal of epidemiology}, 185\penalty0 (10):\penalty0
  950--953, 2017.

\bibitem[Newey(1990)]{newey1990semiparametric}
W.~K. Newey.
\newblock Semiparametric efficiency bounds.
\newblock \emph{Journal of applied econometrics}, 5\penalty0 (2):\penalty0
  99--135, 1990.

\bibitem[Newey and Robins(2017)]{newey2017cross}
W.~K. Newey and J.~M. Robins.
\newblock Cross-fitting and fast remainder rates for semiparametric estimation.
\newblock 2017.

\bibitem[Pearl(2001)]{pearl2001direct}
J.~Pearl.
\newblock Direct and indirect effects.
\newblock In \emph{Proceedings of the seventeenth conference on uncertainty in
  artificial intelligence}, pages 411--420. Morgan Kaufmann Publishers Inc.,
  2001.

\bibitem[Pearl(2009)]{pearl2009causality}
J.~Pearl.
\newblock \emph{Causality}.
\newblock Cambridge university press, 2009.

\bibitem[Pearl(2012)]{pearl2012causal}
J.~Pearl.
\newblock The causal mediation formula—a guide to the assessment of pathways
  and mechanisms.
\newblock \emph{Prevention Science}, 13\penalty0 (4):\penalty0 426--436, 2012.

\bibitem[Robins(1986)]{robins1986new}
J.~Robins.
\newblock A new approach to causal inference in mortality studies with a
  sustained exposure period—application to control of the healthy worker
  survivor effect.
\newblock \emph{Mathematical Modelling}, 7\penalty0 (9-12):\penalty0
  1393--1512, 1986.

\bibitem[Robins et~al.(2017)Robins, Li, Mukherjee, Tchetgen~Tchetgen, and
  van~der Vaart]{robins2017higher}
J.~Robins, L.~Li, R.~Mukherjee, E.~Tchetgen~Tchetgen, and A.~van~der Vaart.
\newblock Higher order estimating equations for high-dimensional models.
\newblock \emph{Annals of statistics}, 45\penalty0 (5):\penalty0 1951, 2017.

\bibitem[Robins et~al.(2000)Robins, Rotnitzky, and
  Scharfstein]{robins2000sensitivity}
J.~M. Robins, A.~Rotnitzky, and D.~O. Scharfstein.
\newblock Sensitivity analysis for selection bias and unmeasured confounding in
  missing data and causal inference models.
\newblock In \emph{Statistical models in epidemiology, the environment, and
  clinical trials}, pages 1--94. Springer, 2000.

\bibitem[Shpitser(2013)]{shpitser2013counterfactual}
I.~Shpitser.
\newblock Counterfactual graphical models for longitudinal mediation analysis
  with unobserved confounding.
\newblock \emph{Cognitive science}, 37\penalty0 (6):\penalty0 1011--1035, 2013.

\bibitem[Sj{\"o}lander(2018)]{sjolander7mediation}
A.~Sj{\"o}lander.
\newblock Mediation analysis with attributable fractions.
\newblock \emph{Epidemiologic Methods}, 7\penalty0 (1), 2018.

\bibitem[Sj{\"o}lander and Vansteelandt(2010)]{sjolander2010doubly}
A.~Sj{\"o}lander and S.~Vansteelandt.
\newblock Doubly robust estimation of attributable fractions.
\newblock \emph{Biostatistics}, 12\penalty0 (1):\penalty0 112--121, 2010.

\bibitem[Tchetgen~Tchetgen and Shpitser(2012)]{tchetgen2012semiparametric}
E.~J. Tchetgen~Tchetgen and I.~Shpitser.
\newblock Semiparametric theory for causal mediation analysis: efficiency
  bounds, multiple robustness, and sensitivity analysis.
\newblock \emph{Annals of statistics}, 40\penalty0 (3):\penalty0 1816, 2012.

\bibitem[Tchetgen~Tchetgen and Shpitser(2014)]{tchetgen2014estimation}
E.~J. Tchetgen~Tchetgen and I.~Shpitser.
\newblock Estimation of a semiparametric natural direct effect model
  incorporating baseline covariates.
\newblock \emph{Biometrika}, 101\penalty0 (4):\penalty0 849--864, 2014.

\bibitem[Van~der Laan and Rose(2011)]{van2011targeted}
M.~J. Van~der Laan and S.~Rose.
\newblock \emph{Targeted learning: causal inference for observational and
  experimental data}.
\newblock Springer Science \& Business Media, 2011.

\bibitem[VanderWeele and Arah(2011)]{vanderweele2011bias}
T.~J. VanderWeele and O.~A. Arah.
\newblock Bias formulas for sensitivity analysis of unmeasured confounding for
  general outcomes, treatments, and confounders.
\newblock \emph{Epidemiology (Cambridge, Mass.)}, 22\penalty0 (1):\penalty0
  42--52, 2011.

\bibitem[Vansteelandt and VanderWeele(2012)]{vansteelandt2012natural}
S.~Vansteelandt and T.~J. VanderWeele.
\newblock Natural direct and indirect effects on the exposed: effect
  decomposition under weaker assumptions.
\newblock \emph{Biometrics}, 68\penalty0 (4):\penalty0 1019--1027, 2012.

\end{thebibliography}


\begin{thebibliography}{5}
\providecommand{\natexlab}[1]{#1}
\providecommand{\url}[1]{\texttt{#1}}
\expandafter\ifx\csname urlstyle\endcsname\relax
  \providecommand{\doi}[1]{doi: #1}\else
  \providecommand{\doi}{doi: \begingroup \urlstyle{rm}\Url}\fi

\bibitem[Pearl(2009)]{pearl2009causality}
J.~Pearl.
\newblock \emph{Causality}.
\newblock Cambridge university press, 2009.

\bibitem[Robins et~al.(1992)Robins, Mark, and Newey]{robins1992estimating}
J.~M. Robins, S.~D. Mark, and W.~K. Newey.
\newblock Estimating exposure effects by modelling the expectation of exposure
  conditional on confounders.
\newblock \emph{Biometrics}, pages 479--495, 1992.

\bibitem[Shpitser and Tchetgen~Tchetgen(2016)]{shpitser2016causal}
I.~Shpitser and E.~J. Tchetgen~Tchetgen.
\newblock Causal inference with a graphical hierarchy of interventions.
\newblock \emph{The Annals of Statistics}, 44\penalty0 (6):\penalty0
  2433--2466, 2016.

\bibitem[Sj{\"o}lander(2018)]{sjolander7mediation}
A.~Sj{\"o}lander.
\newblock Mediation analysis with attributable fractions.
\newblock \emph{Epidemiologic Methods}, 7\penalty0 (1), 2018.

\bibitem[Vansteelandt and VanderWeele(2012)]{vansteelandt2012natural}
S.~Vansteelandt and T.~J. VanderWeele.
\newblock Natural direct and indirect effects on the exposed: effect
  decomposition under weaker assumptions.
\newblock \emph{Biometrics}, 68\penalty0 (4):\penalty0 1019--1027, 2012.

\end{thebibliography}
\end{document}


\onehalfspacing
	\maketitle
	
\section{Proofs of lemmas and theorems} 	
	
	\subsection{Proof of Lemma 1. Generalized front-door functional derivation}
	
	\begin{align*} 
	\Psi & = E[Y(Z(a^*))] \\
	& = \sum_{c,a,z} E(Y(a,z)| Z(a^*)=z,A=a,C=c) Pr(Z(a^*) = z | A=a, C=c) Pr(A=a,C=c) \\
	& \stackrel{M2}{=} \sum_{c,a,z} E(Y(a,z)| Z(a^*)=z,A=a,C=c) Pr(Z(a^*) = z | A=a^*, C=c) Pr(A=a,C=c) \\
	& \stackrel{M1,M3}{=} \sum_{c,a,z} E(Y(a,z)| A=a,C=c) Pr(Z = z | A=a^*, C=c) Pr(A=a,C=c) \\
	& \stackrel{M3}{=} \sum_{c,a,z} E(Y(a,z)| Z=z,A=a,C=c) Pr(Z = z | A=a^*, C=c) Pr(A=a,C=c)\\
	& \stackrel{M1}{=} \sum_{c,a,z} E(Y| Z=z,A=a,C=c) Pr(Z = z | A=a^*, C=c) Pr(A=a,C=c)\\
	& = \sum_{z,c}Pr(Z = z | A=a^*, C=c) \sum_a E(Y| Z=z,A=a,C=c)Pr(A=a | C=c) Pr(C=c)
	\end{align*}

	\subsection{Proof of Lemma 2.}
	
	\begin{align*}
	E[Y \frac{f(Z | a^*, C)}{f(Z | A, C)}] & = \sum_{y,a,z,c} y \frac{f(z | a, c)}{f(z | a, c)} f(y | a,z,c) f(z | a^*, C) f(a | c) f(c) \\ 
	& = \sum_{a,z,c} E(Y | a, z,c) f(z | a^*, c) f(a | c) f(c) \\ 
	& = \Psi
	\end{align*}
	
	The proof of asymptotic normality is fairly standard under the usual regularity conditions once unbiasedness of the estimating equation is established (see Theorem 1A in \cite{robins1992estimating}).
	
	\subsection{Proof of Lemma 3.} 
	
	\begin{align*}
	E\bigg[ \frac{I(A = a^*)}{f(A | C)} E\big( E \big\{ Y | A,Z,C  \big\} | C \big) \bigg] & = \sum_{z,\bar{a},c} I(\bar{a}= a^*) f(z | \bar{a}, c) f(c) E\big( E \big\{ Y | A,z,c  \big\} | c \big) \\
	& = \sum_{z,c} f(z | a^*, c) f(c) \sum_{a} E (Y | a,z,c) f(a | c)  \\
	& = \Psi
	\end{align*}
	
	The proof of asymptotic normality is fairly standard under the usual regularity conditions once unbiasedness of the estimating equation is established (see Theorem 1A in \cite{robins1992estimating}).
	
	\subsection{Proof of Theorem 1. Efficient influence function derivation}
	
	\noindent We aim to find an efficient influence function, $\varphi^{eff}(Y,Z,A,C)$, for $\Psi = E[Y(Z(a^*))]$ under model corresponding to Figure 1c. Our functional is nonparametrically identified under the causal model represented by a complete graph. In other words, the causal model induces no restrictions on the observed data.  Thus, there is a unique influence function, $\varphi^{eff}(Y,Z,A,C)$, and it achieves the semiparametric efficient bound of $\Psi$ in $\mathcal{M}_{np}$. We will use the definition of pathwise differentiability to find the efficient influence function.
	
	$$\frac{d}{dt} \Psi(F_t) = E[ \varphi^{eff}(Y,Z,A,C) \times S(Y,A,Z,C) ] $$  
	
	\noindent where $S(Y,A,Z,C)$ is the score corresponding to the whole model. 
	
	\begin{align}
	\frac{d}{dt} \Psi(F_t) & = \sum_{z,a,c} \frac{d}{dt} E_t[Y | A = a, Z=z, C=C] f_t(z | a^*, c) f_t(a | c) f_t(c) \notag \\
	& \textrm{(from now on, for convenience I will just use } f \textrm{ instead of } f_t)\notag \\
	& = \sum_{z,a,c} \sum_y y \frac{d}{dt} ( f(y | a, z,c) f(z | a^*, c) f(a | c) f(c) ) \notag \\
	& =  \sum_{z,a,c} \sum_y y S(y | a, z, c) f(y | a,z,c) f(z | a^*, c) f(a | c) f(c) \notag  \\
	& + \sum_{z,a,c} E[Y | a, z, c] S(z | a^*, c) f(z | a^*, c) f(a | c) f(c) \notag \\
	& + \sum_{z,a,c} E[Y | a, z, c] f(z | a^*,c) S(a | c) f(a | c) f(c) \notag \\
	& + \sum_{z,a,c} E[Y | a, z, c] f(z | a^*, c) f(a | c) S(c) f(c) \notag \\
	& \textrm{...detail for each of the four terms portion given below...} \notag \\
	& = E\big[ (Y - E(Y | A,Z,C)) \frac{f(Z | a^*, C)}{f(Z | A,C)} \times S(Y, A, Z, C) \big] \label{term1} \tag{A1} \\
	& + E\bigg[ \bigg( \sum_a E[Y | a, Z, C] f(a | C)  \notag\\
	& \hspace{2cm} - \sum_{a, \bar{m}} E(Y | a, \bar{m}, C) f(\bar{m} | \bar{A}, C) f(a | C) \bigg)  \frac{I(\bar{A} = a^*)}{f(\bar{A} | c)} \times S(Y, A, Z, C) \bigg] \label{term2} \tag{A2} \\
	& + E \bigg[ \bigg( \sum_{m} E[Y | A, z, C] f(z | a^*,C) \notag \\
	& \hspace{2cm} - \sum_{a,m} E[Y | a,z,C] f(z | a^*,C) f(a | C) \bigg)\times S(Y, A, Z, C) \bigg] \label{term3} \tag{A3} \\
	& + E \bigg[ \bigg( \sum_{z,a} E[Y | a,z,C] f(z | a^*, C) f(a | C) - \Psi \bigg) \times S(Y, A, Z, C) \bigg] \label{term4} \tag{A4}
	\end{align}
	
	\noindent Each of the four terms (\ref{term1})-(\ref{term4}) will be handled in turn. The goal is to get them in the form $E[ IF \times S] = \sum_{i=1}^4 E[ IF_{i} \times S(Y,A,Z,C) ]$. 
	
	\begin{align*}
	\textrm{(\ref{term1})} & =  \sum_{z,a,c} \sum_y y S(y | a, z, c) f(y | a,z,c) f(z | a^*, c) f(a | c) f(c) \\
	& = \sum_{z,a,c,y} y \frac{f(z | a^*, c)}{f(z | a,c)} f(z | a, c) f(y | a,z,c) f(a | c) f(c)  S(y | a, z, c) \\
	& \stackrel{*}{=} \sum_{z,a,c,y} (y - E[Y | a,z,c]) \frac{f(z | a^*, c)}{f(z | a,c)} f(y | a,z,c) f(z | a, c)  f(a | c) f(c)  S(y | a, z, c) \\
	& \stackrel{**}{=} \sum_{z,a,c,y} (y - E[Y | a,z,c]) \frac{f(z | a^*, c)}{f(z | a,c)} f(y | a,z,c) f(z | a, c)  f(a | c) f(c) \\
	& \hspace{2cm} \times \big( S(y | a, z, c) + S(z | a,c) + S(a | c) + S(c) \big)  \\
	& = \sum_{z,a,c,y} (y - E[Y | a,z,c]) \frac{f(z | a^*, c)}{f(z | a,c)} S(y, a, z, c) f(y, a, z, c) \\
	& = E\big[ (Y - E(Y | A,Z,C)) \frac{f(Z | a^*, C)}{f(Z | A,C)} \times S(Y, A, Z, C) \big]
	\end{align*}
	
	\noindent * The equality will hold because the added term will evaluate to zero as the expectation of a score is zero (in brackets), \\
	$$\sum_{z,a,c} E[Y | a,z,c] \frac{f(z | a^*, c)}{f(z | a,c)} f(z | a, c)  f(a | c) f(c)  \bigg[ \sum_{y} S(y | a, z, c) f(y | a,z,c) \bigg] = 0$$
	\noindent ** Similar to above, the additional terms will all evaluate to zero as the term in the large brackets is zero: 
	\begin{align*}
	\sum_{z,a,c} \frac{f(z | a^*, c)}{f(z | a,c)} f(z | a, c)  f(a | c) f(c)  S( z | a, c) \bigg[ \sum_{y}  (y - E[Y | a,z,c]) f(y | a,z,c) \bigg] & = 0 \\
	\sum_{z,a,c} \frac{f(z | a^*, c)}{f(z | a,c)} f(z | a, c)  f(a | c) f(c)  S(a | c)\bigg[ \sum_{y}  (y - E[Y | a,z,c]) f(y | a,z,c) \bigg] & = 0 \\
	\sum_{z,a,c} \frac{f(z | a^*, c)}{f(z | a,c)} f(z | a, c)  f(a | c) f(c)  S(c) \bigg[ \sum_{y}   (y - E[Y | a,z,c]) f(y | a,z,c) \bigg] & = 0
	\end{align*}
	
	\begin{align*}
	\textrm{(\ref{term2})}  & =  \sum_{z,a,c} E[Y | a, z, c] S(z | a^*, c) f(z | a^*, c) f(a | c) f(c) \\
	& =  \sum_{z,c} \sum_{\bar{a}} \bigg( \sum_a E[Y | a, z, c] f(a | c) \bigg) I(\bar{a} = a^*)S(z | \bar{a}, c) f(z | \bar{a}, c)  f(c) \\
	& =  \sum_{z,c,\bar{a}} \bigg( \sum_a E[Y | a, z, c] f(a | c) \bigg) \frac{I(\bar{a} = a^*)}{f(\bar{a} | c)} S(z | \bar{a}, c) f(z | \bar{a}, c) f(\bar{a} | c)  f(c) \\
	& \stackrel{*}{=}  \sum_{z,c,\bar{a}} \bigg( \sum_a E[Y | a, z, c] f(a | c)  - \sum_{a, \bar{m}} E(Y | a, \bar{m}, c) f(\bar{m} | \bar{a}, c) f(a | c) \bigg) \\
	& \hspace{2cm} \times \frac{I(\bar{a} = a^*)}{f(\bar{a} | c)} S(z | \bar{a}, c) f(z | \bar{a}, c) f(\bar{a} | c)  f(c) \\
	& \stackrel{**}{=} \sum_{z,c,\bar{a}} \bigg( \sum_a E[Y | a, z, c] f(a | c)  - \sum_{a, \bar{m}} E(Y | a, \bar{m}, c) f(\bar{m} | \bar{a}, c) f(a | c) \bigg) \\
	& \hspace{2cm} \times \frac{I(\bar{a} = a^*)}{f(\bar{a} | c)} [S(z | \bar{a}, c) + S(a | c) + S(c) ] f(z | \bar{a}, c) f(\bar{a} | c)  f(c) \\
	& =  \sum_{z,c, \bar{a}, y}\bigg( \sum_a E[Y | a, z, c] f(a | c)  - \sum_{a, \bar{m}} E(Y | a, \bar{m}, c) f(\bar{m} | \bar{a}, c) f(a | c) \bigg) \\
	& \hspace{2cm} \times \frac{I(\bar{a} = a^*)}{f(\bar{a} | c)} [S(y | z, \bar{a}, c) + S(z | \bar{a}, c) + S(a | c) + S(c) ] f(y | z, \bar{a}, c) f(z | \bar{a}, c) f(\bar{a} | c)  f(c) \\
	& =  \sum_{z,c, \bar{a}, y} \bigg( \sum_a E[Y | a, z, c] f(a | c)  - \sum_{a, \bar{m}} E(Y | a, \bar{m}, c) f(\bar{m} | \bar{a}, c) f(a | c) \bigg) \\
	& \hspace{2cm} \times  \frac{I(\bar{a} = a^*)}{f(\bar{a} | c)}  f(y, \bar{a}, z, c) S(y, a, z, c) \\
	& = E\bigg[ \bigg( \sum_a E[Y | a, Z, C] f(a | C)  - \sum_{a, \bar{m}} E(Y | a, \bar{m}, C) f(\bar{m} | \bar{A}, C) f(a | C) \bigg)  \frac{I(\bar{A} = a^*)}{f(\bar{A} | c)} \times S(Y, A, Z, C) \bigg]
	\end{align*}
	
	\noindent *The reasoning here is identical to that for the first term. \\
	\noindent **The reasoning here is identical to that for the first term.  \\
	
	\begin{align*}
	\textrm{(\ref{term3})} & = \sum_{z,a,c} E[Y | a, z, c] f(z | a^*,c) S(a | c) f(a | c) f(c) \\
	& \stackrel{*}{=}\sum_{c,a} \big( \sum_{m} E[Y | a, z, c] f(z | a^*,c) - \sum_{a,m} E[Y | a,z,c] f(z | a^*,c) f(a | c) \big) S(a | c) f(a | c) f(c) \\
	& \stackrel{**}{=} \sum_{c,a} \sum_{\bar{m},y} \big( \sum_{m} E[Y | a, z, c] f(z | a^*,c) - \sum_{a,m} E[Y | a,z,c] f(z | a^*,c) f(a | c) \big) \\
	& \hspace{2cm} \times f(y | a, \bar{m}, c) f(\bar{m} | a, c) f(a | c) f(c) [S(y | a, \bar{m}, c) + S(\bar{m} | a, c) + S(a | c) + S(c) ] \\
	& = E \bigg[ \bigg( \sum_{m} E[Y | A, z, C] f(z | a^*,C) - \sum_{a,m} E[Y | a,z,C] f(z | a^*,C) f(a | C) \bigg)\times S(Y, A, Z, C) \bigg]
	\end{align*}
	
	\noindent *The reasoning here is identical to that for the first term. \\
	\noindent **The reasoning here is identical to that for the first term.  
	
	\begin{align*}
	\textrm{(\ref{term4})} & = \sum_{z,a,c} E[Y | a, z, c] f(z | a^*, c) f(a | c) S(c) f(c)  \\
	& \stackrel{*}{=} \sum_{c} \bigg( \sum_{z,a} E[Y | a,z,c] f(z | a^*, c) f(a | c) - \sum_{z,a,c} E[Y | a,z,c] f(z | a^*, c) f(a | c) f(c) \bigg) S(c) f(c) \\
	& \stackrel{**}{=} \sum_{c} \sum_{y,\bar{a},\bar{m}} \bigg( \sum_{z,a} E[Y | a,z,c] f(z | a^*, c) f(a | c) - \sum_{z,a,c} E[Y | a,z,c] f(z | a^*, c) f(a | c) f(c) \bigg) \\
	& \hspace{2cm} \times f(y | \bar{a}, \bar{m}, c) f(\bar{m} | \bar{a}, c) f(\bar{a} | c) f(c) [S(y | \bar{a}, \bar{m}, c) + S(\bar{m} | \bar{a}, c) + S(\bar{a} | c) + S(c) ] \\
	& = E \bigg[ \bigg( \sum_{z,a} E[Y | a,z,C] f(z | a^*, C) f(a | C) - \sum_{z,a,c} E[Y | a,z,c] f(z | a^*, c) f(a | c) f(c) \bigg) \times S(Y, A, Z, C) \bigg]  \\
	& = E \bigg[ \bigg( \sum_{z,a} E[Y | a,z,C] f(z | a^*, C) f(a | C) - \Psi \bigg) \times S(Y, A, Z, C) \bigg] 
	\end{align*}
	
	\noindent *The reasoning here is identical to that for the first term. \\
	\noindent **The reasoning here is identical to that for the first term.  \\
	
	\noindent Thus, the efficient influence function under the nonparametric model is as follows: 
	
	\begin{align*} 
	\varphi^{eff}(Y,Z,A,C) & = E \bigg[ (Y - E(Y | A,Z,C)) \frac{f(Z | a^*, C)}{f(Z | A,C)} \\
	& \hspace{1cm} + \frac{I(A = a^*)}{f(A | C)} \big( \sum_a E[Y | a, Z, C] f(a | C)  - \sum_{a, \bar{z}} E(Y | a, \bar{z}, C) f(\bar{z} | A, C) f(a | C) \big)  \\
	& \hspace{2cm} + \sum_{z} E[Y | A, z, C] f(z | a^*,C)  \bigg] - \Psi
	\end{align*}
	
	\subsection{Proof of Theorem 2.}

		 We first show that the influence function derived in Theorem 1 has expectation 0 if one of the following scenarios holds: 
		\begin{enumerate} 
			\item  $E(Y | a,z,c)$ and $f(a | c)$ are correct
			\item $f(z | a,c)$ is correct 
		\end{enumerate}
		
		\begin{align*} 
		& \hspace{-1.4cm} \textrm{\underline{1. $E(Y | a,z,c) \textrm{ \& } f(a | c)$ correctly specified and $\tilde{f}(z | a, c)$ misspecified}} \\
		E[\varphi^{eff}] & = E[(Y - E(Y | A,Z,C)) \frac{\tilde{f}(Z | a^*, C)}{\tilde{f}(Z | A,C)}] \\
		& \hspace{1cm} + E[\frac{I(A = a^*)}{f(A | C)} \big( \sum_a E[Y | a, Z, C] f(a | C)  - \sum_{a, z} E(Y | a, z, C) \tilde{f}(z | A, C) f(a | C) \big)] \\
		& \hspace{2cm}+ E[\sum_{z} E[Y | A, z, C] \tilde{f}(z | a^*,C)  - \Psi] \\
		& = 0 + \sum_{a',z,c} \frac{I(a' = a^*)}{f(a' | c)} \bigg( \sum_a E[Y | a, z, c] f(a | c)-  \sum_{a, z} E(Y | a, z, C) \tilde{f}(z | a', c) f(a | c) \bigg) f(z,a',c) \\
		& \hspace{2cm} + E[\sum_{z} E[Y | A, z, C] \tilde{f}(z | a^*,C)  - \Psi] \\
		& = \sum_{z,c} \sum_a E[Y | a, z, c] f(a | c) f(z | a^*,c) f(c) -  \sum_{a, z,c} E(Y | a, z, C) \tilde{f}(z | a^*, c) f(a | c) f(c) \\
		& \hspace{1cm} + E[\sum_{z} E[Y | A, z, C] \tilde{f}(z | a^*,C)  - \Psi] \\
		& = \Psi -  \sum_{a, z,c} E(Y | a, z, C) \tilde{f}(z | a^*, c) f(a | c) f(c)  +  \sum_{a,c} \sum_{z} E[Y | a, z, c] \tilde{f}(z | a^*,c) f(a | c) f(c) - \Psi \\
		& = 0 
		\end{align*} 
		
		\begin{align*} 
		& \hspace{-1.4cm} \textrm{\underline{2. $f(z | a, c)$ correctly specified and $\tilde{E}(Y | a,z,c) \textrm{ \& } \tilde{f}(a | c)$ misspecified}} \\
		E[\varphi^{eff}] & = E[(Y - \tilde{E}(Y | A,Z,C)) \frac{f(Z | a^*, C)}{f(Z | A,C)}] \\
		& \hspace{1cm} + E[\frac{I(A = a^*)}{\tilde{f}(A | C)} \big( \sum_a \tilde{E}[Y | a, Z, C] \tilde{f}(a | C)  - \sum_{a, z} \tilde{E}(Y | a, z, C) f(z | A, C) \tilde{f}(a | C) \big)] \\
		& \hspace{2cm} + E[\sum_{z} \tilde{E}[Y | A, z, C] f(z | a^*,C)  - \Psi] \\
		& = \sum_{c,a,z} (E(Y | a,z,c) - \tilde{E}(Y | a,z,c)) f(z | a^*, c) f(a | c) f(c) \\
		& \hspace{1cm} + \sum_{c,a,z} \frac{1}{\tilde{f}(a | c)} \bigg( \tilde{E}[Y | a, z, c] \tilde{f}(a | c) f(z | a^*,c) f(a^* | c)f(c) \\
		&\hspace{3cm} -  \tilde{E}[Y | a, z, c] \tilde{f}(a | c) f(z | a^*,c) f(a^* | c)f(c) \bigg) \\
		& \hspace{2cm} + \sum_{c,a,z} \tilde{E}[Y | a, z, c] f(z | a^*,c) f(a | c) f(c)  - \Psi \\
		& = \Psi - \sum_{c,a,z} \tilde{E}(Y | a,z,c) f(z | a^*, c) f(a | c) f(c) + \sum_{c,a,z} \tilde{E}[Y | a, z, c] f(z | a^*,c) f(a | c) f(c)  - \Psi \\
		& = 0 
		\end{align*}

		\noindent Assuming the regularity conditions of Theorem 1A in \cite{robins1992estimating} hold for $\varphi^{eff}(Y,Z,A,C)$, the expression follows by standard Taylor expansion arguments: 
	
		$$\sqrt{n} (\hat{\Psi}_{dr} - \Psi) = \frac{1}{\sqrt{n}} \sum_{i=1}^{n} \varphi^{eff}(Y_i, Z_i, A_i, C_i)+ o_p(1) $$
		
		\noindent The asymptotic distribution of the left hand side under $\mathcal{M}_{union}$ follows from the previous equation by the Central Limit Theorem and Slutsky's.

\section{Additional materials}

\subsection{Judea Pearl's front-door criterion} 

\begin{align*}
E(Y(a^*)) & = \sum_{z,c} E(Y(a^*) | Z(a^*) = z, C=c) Pr(Z(a^*) =z | C=c) Pr(C=c) \\
& \stackrel{M2}{=} \sum_{z,c} E(Y(a^*) | Z(a^*) = z, C=c) Pr(Z(a^*) =z | C=c, A=a^*) Pr(C=c) \\
& \stackrel{M1}{=} \sum_{z,c} E(Y(a^*,z) | Z(a^*) = z, C=c) Pr(Z = z | C=c, A=a^*) Pr(C=c) \\
& \stackrel{F1}{=} \sum_{z,c} E(Y(z) | Z(a^*) = z, C=c) Pr(Z = z | C=c, A=a^*) Pr(C=c) \\
& = \sum_{z,c} \bigg[ \sum_a E(Y(z) | Z(a^*) = z, C=c, A=a) Pr(A=a | C=c) \bigg] Pr(Z = z | C=c, A=a^*) Pr(C=c) \\
& \stackrel{M3}{=}  \sum_{z,c} \bigg[ \sum_a E(Y(z) | A=a, C=c) Pr(A=a | C=c) \bigg] Pr(Z = z | C=c, A=a^*) Pr(C=c) \\
& = \sum_{z,c}  Pr(Z = z | C=c, A=a^*) \sum_a E(Y(z) | A=a, C=c) Pr(A=a | C=c) Pr(C=c)\\
& \stackrel{M3}{=} \sum_{z,c}  Pr(Z = z | C=c, A=a^*) \sum_a E(Y(z) | A=a, Z=z, C=c) Pr(A=a | C=c) Pr(C=c)\\
& \stackrel{M1}{=} \sum_{z,c}  Pr(Z = z | C=c, A=a^*) \sum_a E(Y | A=a, Z=z, C=c) Pr(A=a | C=c) Pr(C=c)\\
& = \Psi
\end{align*}

\noindent M3 encodes a so-called ``cross-world" assumption as it posits that $Y(a,z)$ is independent of $Z(a^*)$ given $A$ and $C$, which occur in different worlds (e.g. they cannot be represented in a single world intervention graph). Note that under F1, assumption M3 is no longer a cross-world assumption as it becomes $Y(z) \perp  Z(a^*) \mid A=a,C=c \ \forall \ z,a,a^*,c$ \citep{pearl2009causality}. Additionally, if you happen to evaluate this reduced independence at $a^* = a$, then by consistency, $Y(z) \perp  Z \mid A=a,C=c \ \forall \ z,a,c$. Thus, identification occurs under the Finest Fully Randomized Causally Interpretable Structured Tree Graph interpretation of Figure 1b. We also refer the reader to \cite{shpitser2016causal} where they discuss a general result giving cases when identifying functionals for indirect effects (that generalize the PIIE) and total effects (that generalize the front-door causal effect) coincide (see Lemma 7.3).

\subsection{NPSEM-IE Interpretation of the causal diagram}

\noindent We can formalize the conditions for identification of $\Psi$ under Figure 1c or assumptions M1-M4 using a system of equations known as ``Nonparametric Structural Equation Model". We assign a system of equations for each variable as below: 
\begin{center}
	\begin{align*}
	U & = g_U(\varepsilon_U) \\
	C & = g_C(\varepsilon_C) \\
	A & = g_A(C,U,\varepsilon_A) \\
	Z & = g_Z(A,C,\varepsilon_Z) \\
	Y & = g_Y(Z,A,U,C,\varepsilon_Y) 
	\end{align*}
\end{center}
\noindent Each of the five random variables on this graph are associated with a distinct, arbitrary function, denoted $g$, and a distinct random disturbance, denoted $\varepsilon$, each with a subscript corresponding to its respective random variable. Each variable is generated by its corresponding function, which depends only on all variables that affect it directly. These equations provide a nonparametric algebraic interpretation of the Figure (1c), and are helpful in defining potential outcomes. The identification conditions given above can be formalized in terms of independence conditions about the errors; specifically, we require all the errors to be independent. 

\subsection{Parametric derivation for PIIE}

\noindent Here, we build a parametric expression $E[Y(Z(a^*))]$ where we include parametric models for both $Y$ and $Z$ and leave the joint distribution of $A,C$ unspecified (e.g. estimated by its empirical distribution as described for $\hat{\Psi}^{alt}_{mle}$). We will compare the parametric estimator $\hat{\Psi}^{alt}_{mle}$ to the semiparametric estimators $\hat{\Psi}_{1},\hat{\Psi}_{2},\hat{\Psi}_{dr}$  in our simulation study. The two models are as follows: 

$$ E[Y | A=a, Z=z, C=c] = \theta_0 + \theta_1 a + \theta_2 z + \theta_3 az + \theta_4^T c $$
$$ E[Z | A=a^*, C=c] = \beta_0 + \beta_1a^* + \beta_2^Tc $$

\begin{align*}
E[Y(Z(a^*))] & = \sum_{c} \sum_{z} Pr(Z = z | A=a^*, C=c)  \sum_{a} E(Y | A=a, Z=z, C=c) Pr(A=a, C=c) \\
& =  \sum_{c} \sum_{z} Pr(Z = z | A=a^*, C=c)  \sum_{a} (\theta_0 + \theta_1 a + \theta_2 z + \theta_3 az + \theta_4^T c) Pr(A=a,C=c)\\
& = \sum_{c} Pr(C=c) \sum_{z} Pr(Z = z | A=a^*, C=c)  \times \\
& \hspace{3.5cm}\big(\theta_0 + \theta_1 E[A | C=c] + \theta_2 z + \theta_3z E[A | C=c] + \theta_4^T c \big) \\
& = \theta_0 + \theta_1 \sum_{c} E[A | C=c] Pr(C=c) + \theta_2  \sum_{c} Pr(C=c)  \sum_{z} z Pr(Z = z | A=a^*, C=c)  \\
& \hspace{.7cm} + \theta_3  \sum_{c} Pr(C=c) \sum_{z} z E[A | C=c] Pr(Z = z | A=a^*, C=c) + \theta_4^T E[C] \\
& = \theta_0 + \theta_1 E[A] + \theta_2  \sum_{c} Pr(C=c)  E[Z | A=a^*, C=c]  \\
& \hspace{2cm} + \theta_3  \sum_{c} Pr(C=c) E[A | C=c] E[Z | A=a^*, C=c] + \theta_4^T E[C] \\
& =  \theta_0 + \theta_1 E[A] + \theta_2  \sum_{c} Pr(C=c)  (\beta_0 + \beta_1a^* + \beta_2^Tc)  \\
& \hspace{2cm} + \theta_3  \sum_{c} Pr(C=c) E[A | C=c] (\beta_0 + \beta_1a^* + \beta_2^Tc) + \theta_4^T E[C] \\
& =  \theta_0 + \theta_1 E[A] + \theta_2 \beta_0 + \theta_2 \beta_1 a^* + \theta_2 \beta_2^T E[C] + \theta_3 \beta_0 E[A] + \theta_3 \beta_1 a^* E[A] \\
& \hspace{2cm} + \theta_3 \beta_2^T  \sum_{c} c Pr(C=c) E[A | C=c]+ \theta_4^T E[C] \\
& =  \theta_0 + \theta_1 E[A] + \theta_2 \beta_0 + \theta_2 \beta_1 a^* + \theta_2 \beta_2^T E[C] + \theta_3 \beta_0 E[A] + \theta_3 \beta_1 a^* E[A] \\
& \hspace{2cm} + \theta_3 \beta_2^T  E[C E[A | C=c]]+ \theta_4^T E[C] \\
& = \theta_0 + \theta_2 \beta_0 + \theta_2 \beta_1 a^* + (\theta_1 + \theta_3 \beta_0 + \theta_3 \beta_1 a^*)E[A] + (\theta_2 \beta_2^T + \theta_4^T) E[C] + \theta_3 \beta_2^T  E[AC]
\end{align*}

\noindent For estimation, the empirical mean can be used for $E[A]$, $E[C]$, and $E[AC]$. In this setting, there is a closed form expression for the variance. In settings where the outcome or mediator variable are binary, the variance can be computed using the sandwich variance or via the nonparametric bootstrap.

\begin{align*} 
 Var(PIIE) & =  \beta_1 \theta_2(\beta_1 \theta_3 Cov(A,A^2) +  \beta_1 \theta_2Var(A)) + \beta_1 \theta_3 (Var(A^2) \beta_1 \theta_3 +  \beta_1 \theta_2 Cov(A,A^2)) \\
& \hspace{1cm} + ( E(A) \theta_2 + E(A^2) \theta_3)^2 Var(\beta_1)  + E(A) \beta_1 ( E(A) \beta_1 Var(\theta_2)  + E(A^2) \beta_1 Cov(\theta_2,\theta_3)) \\
& \hspace{2cm} + E(A^2) \beta_1 ( E(A) \beta_1 Cov(\theta_2,\theta_3) + E(A^2) \beta_1 Var(\theta_3) ) 
\end{align*}

\noindent For estimation with binary $A$, $E(A) = E(A^2) = \overline{A}$, $Var(A) = Var(A^2) = Cov(A,A^2) = S_A^2$ (sample variance), and all the parameters are estimated via their MLE in R.

\subsection{Sandwich variance}

\noindent Let $\theta$ denote the vector of all $K$ parameters and $U(\theta) = [U_1^T, ... , U_K^T]^T$ denote the score vector where the $K$th score corresponds to the score for $\Psi$. A consistent estimator for the asymptotic variance of $\theta$:  

$$ \widehat{Var(\theta)} = [\sum_{i=1}^n \frac{dU(\theta)}{d\theta} |_{\theta = \hat{\theta}} ]^{-1} U(\hat{\theta})^T U(\hat{\theta}) [\sum_{i=1}^n \frac{dU(\theta)}{d\theta} |_{\theta = \hat{\theta}} ]^{-1T} $$

\noindent Further, a consistent estimator for the asymptotic variance of $\hat{\Psi}$ will correspond to the $\widehat{Var(\theta)}_{k,k}$ element.  \\

\subsection{Population intervention effect and the total effect among exposed}

For binary $A$ and $a^* = 0$, 

\begin{align*} 
PIE(0) & = E(Y - Y(0)) \\
& = E(AY + (1-A)Y -Y(0)) \\
& = E(AY(1) + (1-A)Y(0) - Y(0)) \textrm{     by Consistency} \\
& = E(A(Y(1) - Y(0)) + Y(0) - Y(0) ) \\
& = E(A(Y(1) - Y(0)) ) \\
& = E( E(A (Y(1) - Y(0)) | A) )  \\
& = E( Y(1) - Y(0) |  A=1) )  Pr(A=1) \\
& = ETT \ Pr(A=1) 
\end{align*}

\subsection{Alternative population intervention effect decomposition} 

We could have used an alternative decomposition of the population intervention effect, 
\begin{align}
PIE(a^*) & = E[Y - Y(a^*,Z(a^*))] = \underbrace{E[Y(a^*,Z) - Y(a^*)]}_\text{ indirect effect} + \underbrace{E[ Y - Y(a^*,Z) ]}_\text{ direct effect} \label{alt_decomp} \tag{A5}
\end{align}
The use of this alternative decomposition would not guarantee robustness to exposure-outcome confounding as the indirect effect includes the term $E(Y(a^*))$, which requires no unmeasured confounding of exposure-outcome relation for identification (similar to our PIDE). Additionally, identification of the term $E(Y(a^*,Z))$ requires a different set of assumptions that will not lead to a connection with the frontdoor formula. Under certain conditions, the indirect and direct effects from (\ref{alt_decomp}) connect to work by \cite{vansteelandt2012natural} and \cite{sjolander7mediation}. We discuss both below. 

\subsubsection{Connection to Vansteelandt and Vanderweele (2012)}

If we were to condition on the exposed, the indirect and direct effects from (\ref{alt_decomp}) aligns with the effect decomposition of the effect of treatment on the treated (ETT), also known as the total effect on the exposed, described by \cite{vansteelandt2012natural},

$$ ETT = E[ Y(1) - Y(0) |  A=1 ]  = \underbrace{E[Y(0,Z) - Y(0) | A = 1]}_\text{natural indirect effect on the exposed} + \underbrace{E[ Y - Y(0,Z) | A = 1]}_\text{natural direct effect on the exposed} $$

\noindent The identification conditions needed for the \cite{vansteelandt2012natural}'s indirect effect are different than those needed for the PIIE. These are listed in their paper (section 4), but we also state them using our notation below: 

\begin{align*} 
\textrm{M1.} \ & \textrm{Consistency assumptions: } \textrm{(1) If $A=a$, then $Z(a) =Z$ w.p.1}, \\
& \hspace{4.5cm} \textrm{(2) If $A=a$, then $Y(a) =Y$ w.p.1}, \\
& \hspace{4.5cm} \textrm{(3) If $A=a$ and $Z=z$, then $Y(a,z) =Y$ w.p.1} \\
\textrm{M3.} \ & Y(a,z) \perp  Z | A=a,C=c \ \  \forall \ z,a,c \\
\textrm{M4.} \ & Y(a,z) \perp  A | C=c \ \ \ \forall \ z,a, c \\
\end{align*}

\noindent These assumptions could be formulated under a Nonparametric Structural Equation Model with Independent Errors (NPSEM-IE) interpretation of the diagram in Figure 1a. Note that they do not follow from an FFRCISTG interpretation of the diagram.

\subsubsection{Connection to Sj{\"o}lander (2018)}

If we were to scale by proportion of persons with outcome, the indirect and direct effects from (\ref{alt_decomp}) aligns with the effect decomposition of the attributable fraction (AF) described by \cite{sjolander7mediation},

$$ AF = E[Y-Y(a^*)]/E[Y]  = \underbrace{E[ Y(a^*,Z)-Y(a^*) ]/E[Y]}_\text{natural indirect attributable fraction} + \underbrace{E[Y-Y(a^*,Z)]/E[Y]}_\text{natural direct attributable fraction} $$

\noindent The identification conditions needed for Sj{\"o}lander (2018)'s indirect effect are the same as listed for \cite{vansteelandt2012natural}'s indirect effect. However, in addition to a consistency assumption, they state the necessary assumption for identification as $Y(a,Z) \perp A | Z,C$, which is implied by M3 and M4. 

\bibliography{mybib}